\documentclass[preprint,prd,aps,superscriptaddress]{revtex4-1}
\usepackage{hyperref}
\usepackage{bm}
\usepackage{amsmath}
\usepackage{amssymb}
\usepackage{graphicx}
\usepackage{verbatim}

\newcommand{\Li}{\mathrm{Li}}
\newcommand{\lnb}{\mathrm{lnb}}

\begin{document}
\title{Next to leading order QED corrections to the process $ \gamma\gamma\rightarrow\mu^+\mu^-\gamma $ }

\author{Mikhail G.  Kozlov}
\affiliation{Budker Institute of  Nuclear Physics, Novosibirsk, 630090 Russia}
\affiliation{Novosibirsk State University, Novosibirsk, 630090 Russia}
\email{m.g.kozlov@inp.nsk.su}

\thanks{This work is supported by the RFBR grants No. 16-32-60033 and 15-02-09016}

\begin{abstract}
We have calculated one loop quantum electrodynamic corrections to the process $ \gamma\gamma\rightarrow\mu^+\mu^-\gamma $, where all photons are on mass shell and the muon mass is taken into account. The result is obtained in the analytical form and is implemented as functions in the \verb|C| programming language, which can be used to calculate the cross-section, the differential cross section, and to construct generators. We also present numerical results for corrections to the cross section and to the differential cross section.

\end{abstract}

\maketitle

\section{Introduction}

In this paper we consider the $ \gamma\gamma\rightarrow f^+ f^-\gamma $ process in the next-to-leading order (NLO), where the photons are on the mass shell and $f^{\pm}$ are fermions such as $e,\,\mu,\,\tau$. 
We calculated analytically the radiative quantum electrodynamics corrections to the square of the amplitude of the process $ \gamma\gamma\rightarrow f^+f^-\gamma $. The result was obtained as functions of Lorentz invariants and the mass of the fermion $f^{\pm}$. Analytical calculations were performed in the \verb|Wolfram Mathematica| using the \verb|LiteRed| package~\cite{litered} to reduce loop integrals to a set of master integrals. The obtained analytical result was implemented as functions in the C programming language. With the help of them numerical calculations were performed for the scattering cross section and for the differential cross section of the process $\gamma\gamma\rightarrow\mu^+\mu^-\gamma$. Using crossing invariance and analytic continuation of master integrals, we can obtain expressions for the NLO corrections to  squares of amplitude for processes $ f^{\pm}\gamma\rightarrow f^{\pm}\gamma\gamma $ and $ f^+f^-\rightarrow 3\gamma $.

The process $\gamma\gamma\rightarrow f^+f^-\gamma$ can be treated as part of the process $ ee\rightarrow ee\,f^+f^-\gamma $ in the two-photon production channel with small virtualities of the photons. The overview of two-photon processes was given in~\cite{BGMS:1975}. The scattering cross section of the process $ ee\rightarrow ee\,f^+f^-\gamma  $  grows logarithmically with increasing energy. Such processes  can be observed by zero-degree or forward detectors in existing colliders. Also the processes $\gamma\gamma\rightarrow f^+f^-\gamma $ can be observed in the future experiments on photon colliders. Processes $\gamma\gamma\rightarrow \mu^+\mu^-\gamma$ and $\gamma\gamma\rightarrow e^+e^-\gamma$ will give background events for processes with a hadronic final state. For example, for the rare two-photon process $ \gamma\gamma\rightarrow \eta (\eta',\dots)\rightarrow \mu^+\mu^-\gamma (e^+e^-\gamma)$ we need to know corrections to accurately account for the background. 

At colliders with high luminosity, such as the C and B factories, the radiative return method is often used. For processes of the type $ e^+e^-\rightarrow \gamma_{ISR} X$ (ISR is initial state radiation), where $ X\rightarrow f^+f^- $ decays into a pair of fermions, two-photon processes $ e^+e^-\rightarrow e^+e^- f^+f^-\gamma $ (with final $ e^+e^- $ undetected) will produce background events if the $ \gamma_{ISR} $  is not detected. 

Using analytical results for the $ \gamma\gamma\rightarrow f^+f^-\gamma $ process, one can obtain expressions for the square of amplitude of the $ e^+e^-\rightarrow 3\gamma $ process in the NLO via crossing invariance. The process $ e^+e^-\rightarrow 3\gamma $ is the main background for the processes $ e^+e^-\rightarrow \pi^0\gamma $, $ e^+e^-\rightarrow \eta\gamma $ etc. Therefore it is important to know the NLO corrections. On the basis of the result obtained, a event generator will be made for the process $ e^+e^-\rightarrow 3\gamma $ in the next-to-leading order. The same radiative corrections to the process of the orthopositronium decay $ e^+e^-\rightarrow 3\gamma $ were considered in~\cite{pos}. The corrections to the process $ e^+e^-\rightarrow 3\gamma  $ (and other processes obtained by crossing) contain, in addition to standard QED corrections, QCD corrections in the form of light by light scattering diagrams as well. Corrections of this sort at low energies do not have good theoretical predictions, hence the accurate  measurement of this process will provide additional information.

\section{Definitions and result}

\begin{figure}[t]
	\includegraphics[width=10cm]{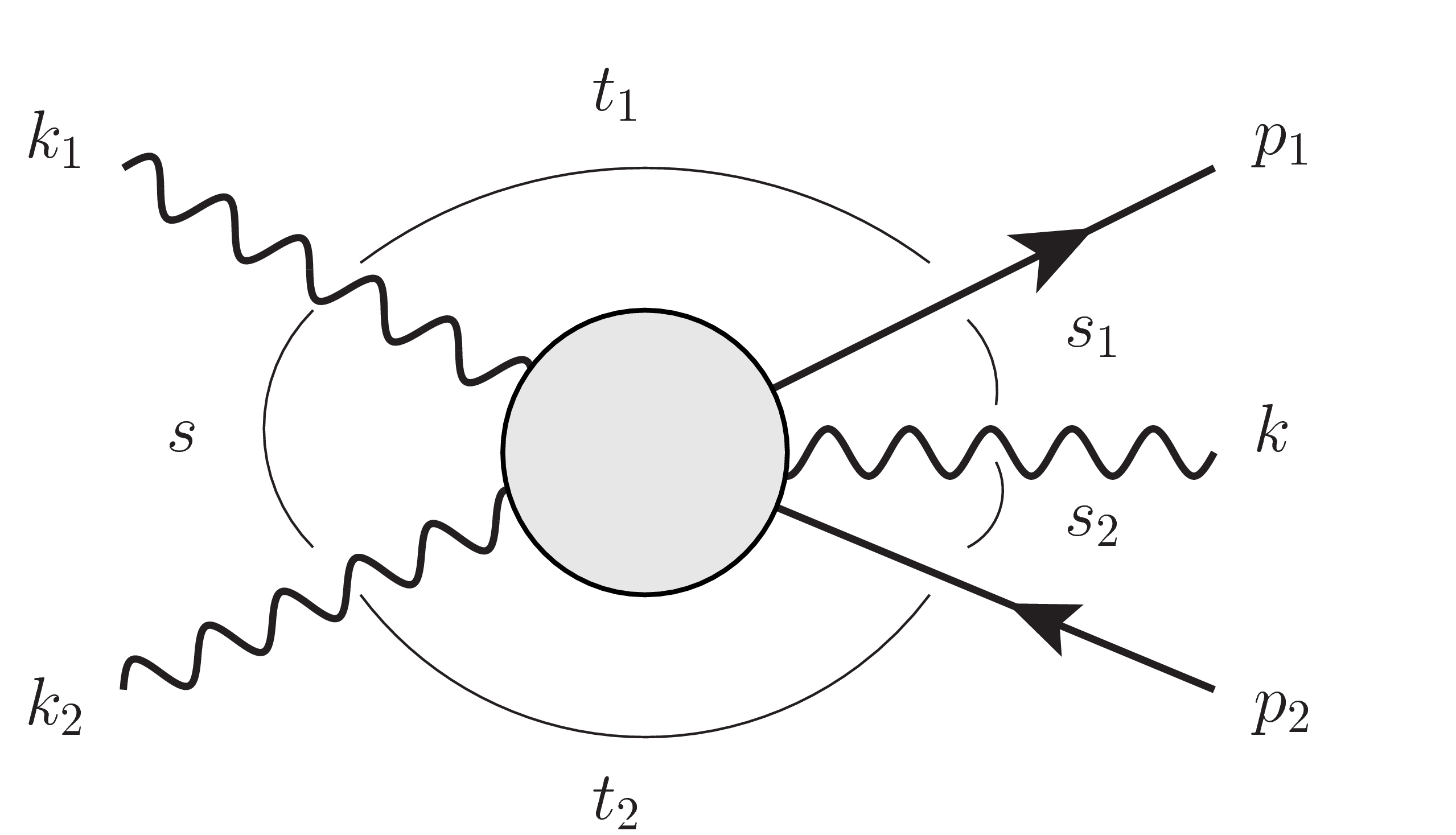}
	\caption{schematic representation of the process $ \gamma\gamma\rightarrow\mu^+\mu^-\gamma. $}\label{fig:1}
\end{figure}

Let us consider the process of $ \gamma\gamma\rightarrow\mu^+\mu^-\gamma $ in quantum electrodynamics (QED) in the next to leading order (NLO) see fig~\ref{fig:1}. We introduce the notation for four-vectors of particles: $ k_1,\,k_2 $ --- are the momenta of the colliding photons, $ p_1,\,p_2$ --- are the momenta of the final muons, and $k $ --- momentum of the final photon. All particles are on the mass shell:
\[
k_1^2=0,\,k_2^2=0,\quad p_1^2=1,\,p_2^2=1,\,k^2=0\,.
\]
We  divide all momenta by muon mass $m_{\mu}$. Here and below we work with dimensionless quantities and the dimension will be reconstructed using the muon mass. Next, we introduce Lorentz invariants
\begin{equation}
s=(k_1+k_2)^2\,,\;s_1=(p_1+k)^2\,,\;s_2=(p_2+k)^2\,,\;t_1=(k_1-p_1)^2\,,\;t_2=(k_2-p_2)^2\,,
\end{equation}
see fig.~\ref{fig:1}. Let us introduce the notation for the amplitude of the process:
\begin{equation}
M_{\gamma\gamma\rightarrow\mu\mu\gamma}=M_{born}+M_{nlo}\,,
\end{equation}
where the first term is the leading order, the second term is the next to leading order amplitude. The square of the amplitude summed over the polarizations of particles reads
\begin{equation}
T_{2\mu\gamma}(\boldsymbol{s})=\frac{1}{4}\sum_{pol}|M_{\gamma\gamma\rightarrow\mu\mu\gamma}|^2\approx \frac{1}{4}\sum_{pol}|M_{born}|^2+\frac{1}{2}\sum_{pol}\Re\bigl(M_{nlo}M_{born}^*\bigr)\,,
\end{equation}
where $ \boldsymbol{s}=(s,s_1,s_2,t_1,t_2) $. The expression for $ \sum_{pol}|M_{\gamma\gamma\rightarrow\mu\mu\gamma}|^2 $ is still quite large and is realized as a function in the \verb|C| programming language~\cite{ggmmg} (see Appendix). Since the quantity $ T_{2\mu\gamma} $ is infrared divergent, to cancel the divergence we need to add a square of the amplitude with the emission of an additional soft photon:
\begin{equation}
\widetilde{T}_{2\mu\gamma}(\boldsymbol{s}) = \frac{1}{4}\sum_{pol}|M_{born}|^2+\frac{1}{2}\sum_{pol}\Re\bigl(M_{nlo}M_{born}^*\bigr)+T_{2\mu 2\gamma}(\boldsymbol{s}, \omega_{\max},n)\,,
\end{equation}
where $ T_{2\mu 2\gamma} $ is  the square of the amplitude  integrated over the momentum of the additional bremsstrahlung photon with maximum energy $ \omega_{\max} $ in the reference frame characterized by a four-vector $ n^{\mu}=(1,0,0,0) $. Four vector $ n^{\mu} $ has normalization $ n^2=1 $. 

In this paper we calculated the following quantity
\begin{equation}\label{eq:T}
T_{2\mu\gamma(nlo)}(\boldsymbol{s}, \omega_{\max},n)=\frac{1}{2}\sum_{pol}\Re\bigl(M_{nlo}M_{born}^*\bigr)+T_{2\mu 2\gamma}(\boldsymbol{s}, \omega_{\max},n)\,,
\end{equation}
as a function of invariants $ s,\,s_1,\,s_2,\,t_1,\,t_2 $, maximum energy of bremsstrahlung photon $ \omega_{\max}$ in reference frame $ n^{\mu} $. The quantity~\eqref{eq:T} is the NLO correction to process $ \gamma\gamma\rightarrow\mu^+\mu^-\gamma $.

The NLO corrections can be divided into three types 
\begin{equation}\label{eq:nlo}
\begin{split}
&T_{2\mu\gamma(nlo)}(\boldsymbol{s},\omega_{\max},n)=T^{(1)}(\boldsymbol{s},\epsilon)+T^{(2)}(\boldsymbol{s},\epsilon)+T^{(3)}(\boldsymbol{s},\epsilon, \omega_{\max},n)\,,\; \epsilon\rightarrow 0,
\end{split}
\end{equation}
where $ d=4-2\epsilon $ is dimension of the space-time in dimensional regularization, $ T^{(1)} $ is the sum of corrections with one fermion line, $ T^{(2)} $ is the sum of corrections with two fermion lines (the amplitude contains a subdiagram of light by light scattering), $ T^{(3)} $ is a real correction from bremsstrahlung photon with maximum energy $ \omega_{\max} $. Expressions $ T^{(1)} $ and $ T^{(3)} $ contain infrared divergences, which cancel in the sum of $ T^{(1)}+T^{(3)} $. Let us extract the divergent parts of the expressions $ T^{(1)} $ and $ T^{(3)} $:
\begin{equation}\label{eq:ir_sing}
\begin{split}
&T^{(1)}(\boldsymbol{s},\epsilon)=-\frac{1}{\epsilon}\Phi(\boldsymbol{s},\epsilon)+T^{(1)}_+(\boldsymbol{s})+{\cal O}(\epsilon)\,,\\
&T^{(3)}(\boldsymbol{s},\epsilon, \omega_{\max},n)=\frac{1}{\epsilon}\Phi(\boldsymbol{s},\epsilon)+T^{(3)}_{+}(\boldsymbol{s}, \omega_{*},n)+T^{(3)}_{2\gamma}(\boldsymbol{s}, \omega_{*},\omega_{\max},n)+{\cal O}(\epsilon)\,,\\
&T^{(2)}(\boldsymbol{s},\epsilon)=T^{(2)}(\boldsymbol{s})+{\cal O}(\epsilon)\,,
\end{split}
\end{equation}
where $ \Phi(\boldsymbol{s},\epsilon) $ is defined in~\eqref{eq:Phi}, the sum $ \Phi(\boldsymbol{s},\epsilon)/\epsilon + T^{(3)}_{+}(\boldsymbol{s}, \omega_{*},n) $ is the real correction in the soft photon approximation with maximum energy of bremsstrahlung photon $ \omega_{*}\ll\omega_{\max} $~, $ T^{(3)}_{2\gamma}(\boldsymbol{s}, \omega_{*},\omega_{\max},n) $ is the real correction with bremsstrahlung photon energies from $ \omega_{*} $ to $ \omega_{\max} $. Hence the NLO corrections take the following form:
\begin{equation}
T_{2\mu\gamma(nlo)}(\boldsymbol{s},\omega_{\max},n)=T^{(1)}_+(\boldsymbol{s})+T^{(2)}(\boldsymbol{s})+T^{(3)}_{+}(\boldsymbol{s}, \omega_{*},n)+T^{(3)}_{2\gamma}(\boldsymbol{s}, \omega_{*},\omega_{\max},n)\,.
\end{equation}

The result of this work is the expression for $ T_+^{(1)} $, $ T^{(2)} $ and $ T_{+}^{(3)} $ implemented in the \verb|C| programming language functions using \verb|GNU Scientific Library (GSL)|~\cite{gsl}. The source code for these functions is available on the website ~\cite{ggmmg}. An example of the use of the functions can be found in the Appendix. 

Corrections to the cross section can be represented as follows
\begin{equation}
\sigma=\sigma_{born}(1+\delta_1+\delta_2+\delta_3)
\end{equation}
where $ \delta_1 $ --- are virtual corrections with one fermion line, $ \delta_2 $  --- are virtual corrections with two fermion lines (with a fermion box diagram) and $ \delta_3 $ --- is a real correction:
\begin{equation}
\begin{split}
&\delta_1=\frac{1}{2s\,\sigma_{born}}\int T^{(1)}_+(\boldsymbol{s})d\Gamma_3\,,\\
&\delta_2=\frac{1}{2s\,\sigma_{born}}\int T^{(2)}(\boldsymbol{s})d\Gamma_3\,,\\
&\delta_3=\frac{1}{2s\,\sigma_{born}}\int \Bigl[T^{(3)}_{s.p.}(\boldsymbol{s}, \omega_{*},n)+T^{(3)}_{2\gamma}(\boldsymbol{s}, \omega_{*},\omega_{\max},n)\Bigr]d\Gamma_3=\\
&\quad=\frac{1}{2s\,\sigma_{born}}\int T^{(3)}_{s.p.}(\boldsymbol{s}, \omega_{*},n)d\Gamma_3+\frac{\sigma_{2\mu 2\gamma}(\omega_*,\omega_{\max},n)}{\sigma_{born}}\,.
\end{split}
\end{equation}
Here $ \sigma_{2\mu 2\gamma} $ is the born cross section of the process $ \gamma\gamma\rightarrow \mu^+\mu^-\gamma\gamma $, where a soft photon has energy in the range from $ \omega_{*} $ to $ \omega_{\max} $ in the reference frame $ n^{\mu} $. The value of $ \sigma_{2\mu 2\gamma} $ can be easily calculated using the package \verb|CompHEP|~\cite{comphep,comphep2,comphep_web}. Expressions for obtaining the scattering cross section are presented in~\eqref{eq:sigma}.

\subsection*{Numerical result for corrections to the cross section}
Here are some of the numerical results obtained using the functions described above.
\begin{figure}
\includegraphics[width=8cm]{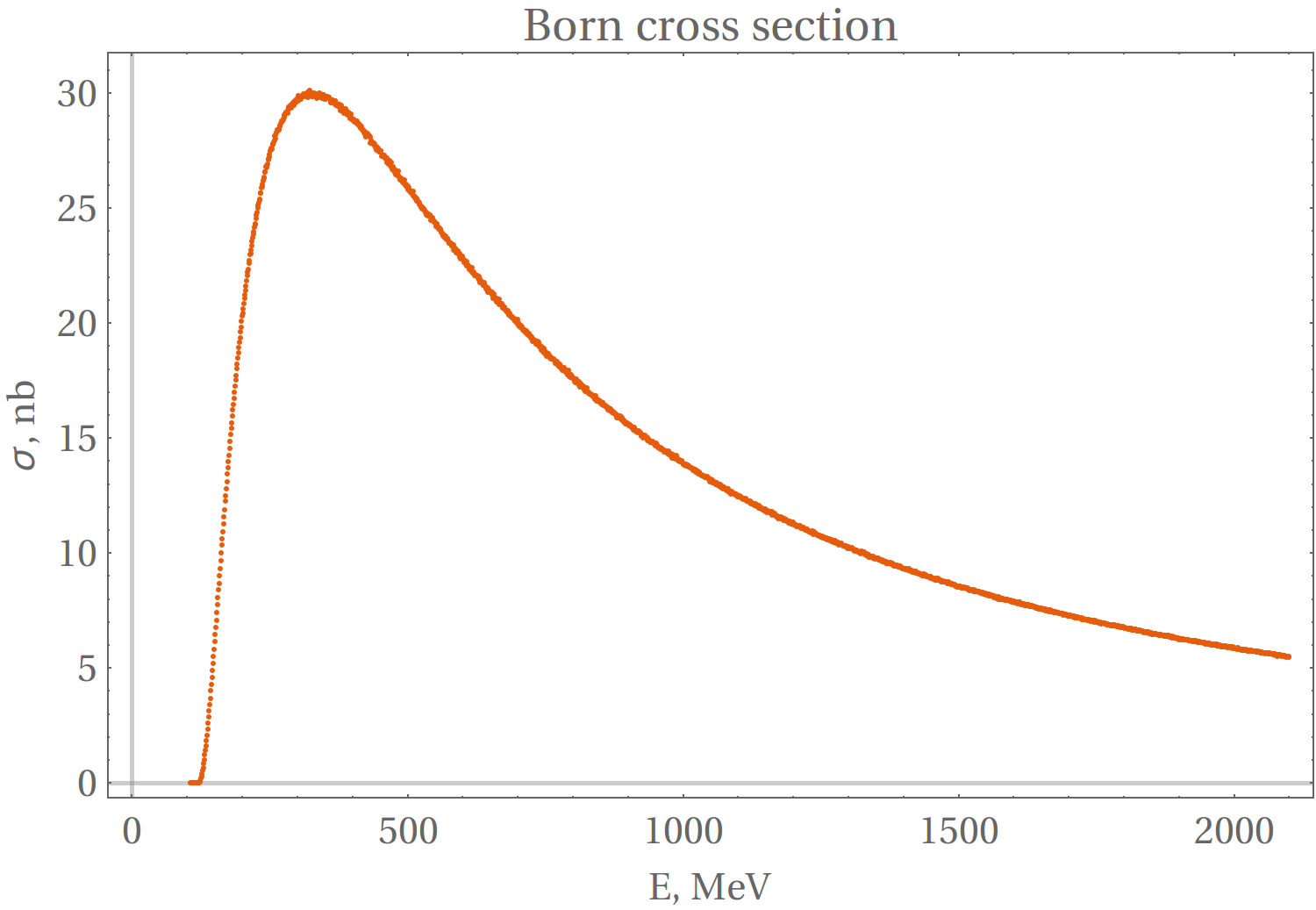}
\includegraphics[width=8cm]{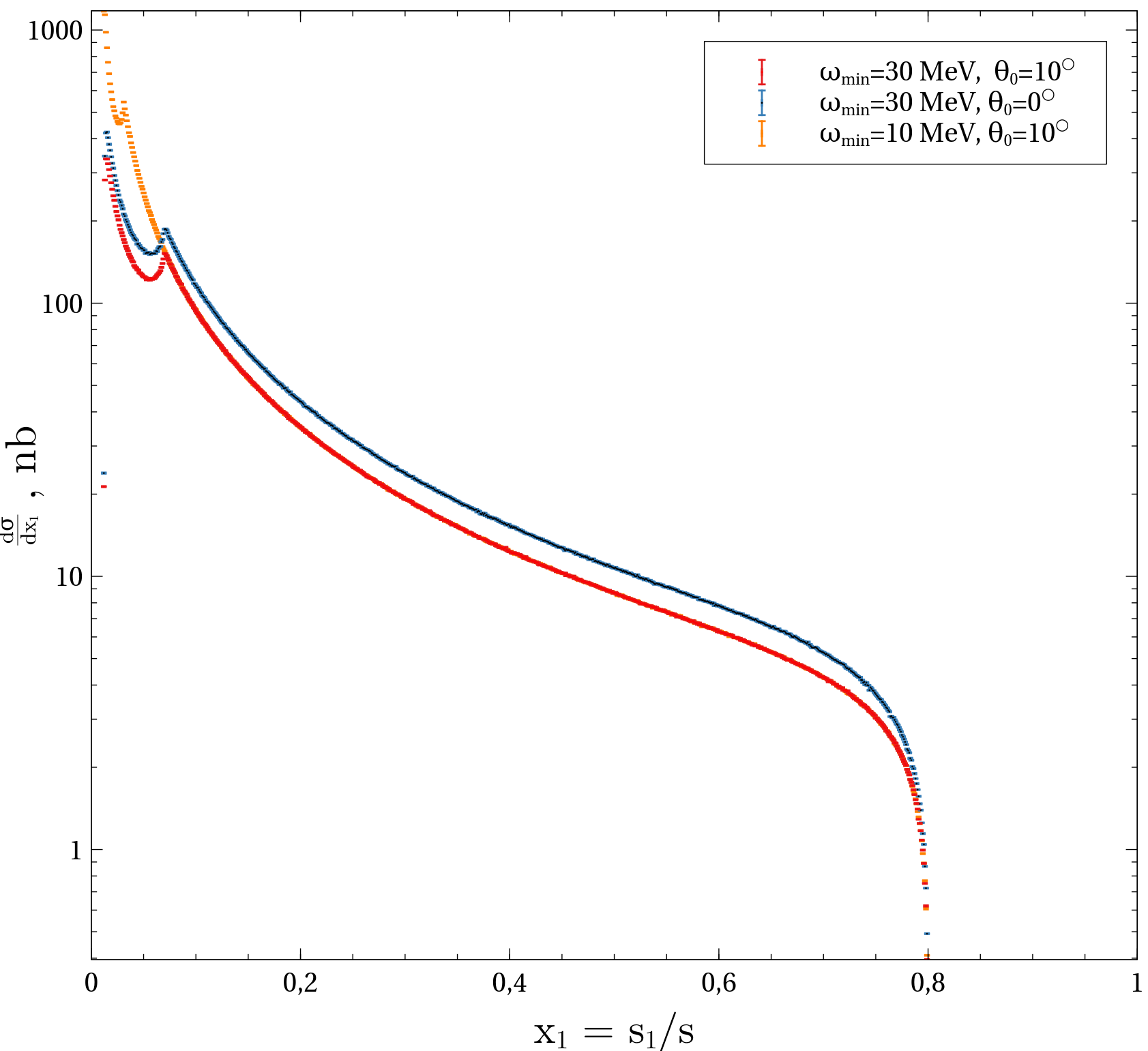}
\caption{Right picture is Born cross section vs the initial photon energy $ \sqrt{s}/2 $. Left picture is differential cross section for Born approximation vs invariant mass of the photon and muon $ s_1=(k+p_1)^2 $. The energy of initial photons is $ \sqrt{s}/2=500 $ MeV. Different graphs correspond to different restrictions on the minimum angle $ \theta_0 $ and the minimum energy $ \omega_{\min} $ of the final photon. The figure shows that the position of a small peak is affected only by the value of the minimum energy $ \omega_{\min} $ of the final photon.}\label{fig:born}
\end{figure}
To obtain numerical results, we used the Monte Carlo method with Vegas algorithm~\cite{vegas}.
We use the following conditions for the final particles when calculating the cross section:
\begin{equation}\label{eq:cuts}
\omega_{min}=30 MeV,\quad \theta_0=10^{\circ}\,,
\end{equation}
where $ \omega_{min} $ --- is the minimum  energy of the final photon, and $ \theta_0$ --- is the minimum angle relative to the beam axis ($ \theta_0 < \theta < \pi-\theta_0 $, where $ \theta $ --- is the particle angle). Figure~\ref{fig:born} shows the dependence of the Born cross section on the photon energy $ \sqrt{s}/2 $ with conditions~\eqref{eq:cuts}.  Numerical results for the corrections are
\begin{center}
\begin{tabular}{c|c|c|c|c|c}
$ \omega_1 $, MeV	& $\sigma_{born}$, nb & $\delta_1$ & $\delta_2$  & $\delta_3$ & $\delta_{2u}$\\ 
\hline
200			& 20.25(1) 			  & 0.0279(5)  & -0.00235(6) & -0.00210(1) & -0.000363(1) \\ 
300			& 29.73(1)			  & 0.031(3)   & -0.00225(5) & -0.0021(1) & -0.000260(1)\\ 
500			& 25.93(1)			  & 0.0277(8)  & -0.00173(5) & -0.0046(2) & -0.000173(1)\\ 
1000		& 13.90(1)			  & 0.015(1)   & -0.00070(5) & -0.0094(2) & -0.000103(1)\\ 
\end{tabular} 
\end{center}
Here $ \omega_1 $ --- is the energy of the initial photon in the center of mass frame ($ k_1=(\omega_1,0,0,\omega_1) $, $ k_2=(\omega_1,0,0,-\omega_1) $), $ \delta_{2u} $ --- is the correction $ \delta_2 $ with only u-quark. The following physical parameters were used in the calculation
\begin{equation}
\begin{split}
&\alpha=1/137.04\,(\text{fine structure constant}) ,\\
&m_{\mu}=105.7\, MeV\,(\text{muon mass}), \\
&m_e=0.511\,MeV\,(\text{electron mass}), \\
&m_{u}=2.0\, MeV\,(\text{u quark mass}) \,.
\end{split}
\end{equation}
Note, that we considered only three contributions in  $ \delta_2 $ : the contribution of the electron loop, the muon loop and the loop of u quark (contribution of $ \tau $ is suppressed by mass).  We made an estimate of the hadron contribution through the contribution of only the u-quark \mbox{($\delta_{2u}/\delta_1\lesssim 0.01$}). The contribution of the remaining quarks is suppressed (if we compare with the electron loop) either by charge (factor $ 1/81 $ for $ d,\,s $) or by mass (for $ c $). 

For real correction $ \delta_3 $ the following parameters were used in the center mass frame:
\begin{equation}
\omega_{*}=0.1\; MeV\,,\quad \omega_{\max}=30\; MeV\,,
\end{equation} 
where $ \omega_* $ is the maximum energy for the soft photon approximation, $ \omega_{\max} $ is maximum energy of the soft photon.

\begin{figure}[t]
	\includegraphics[width=10cm]{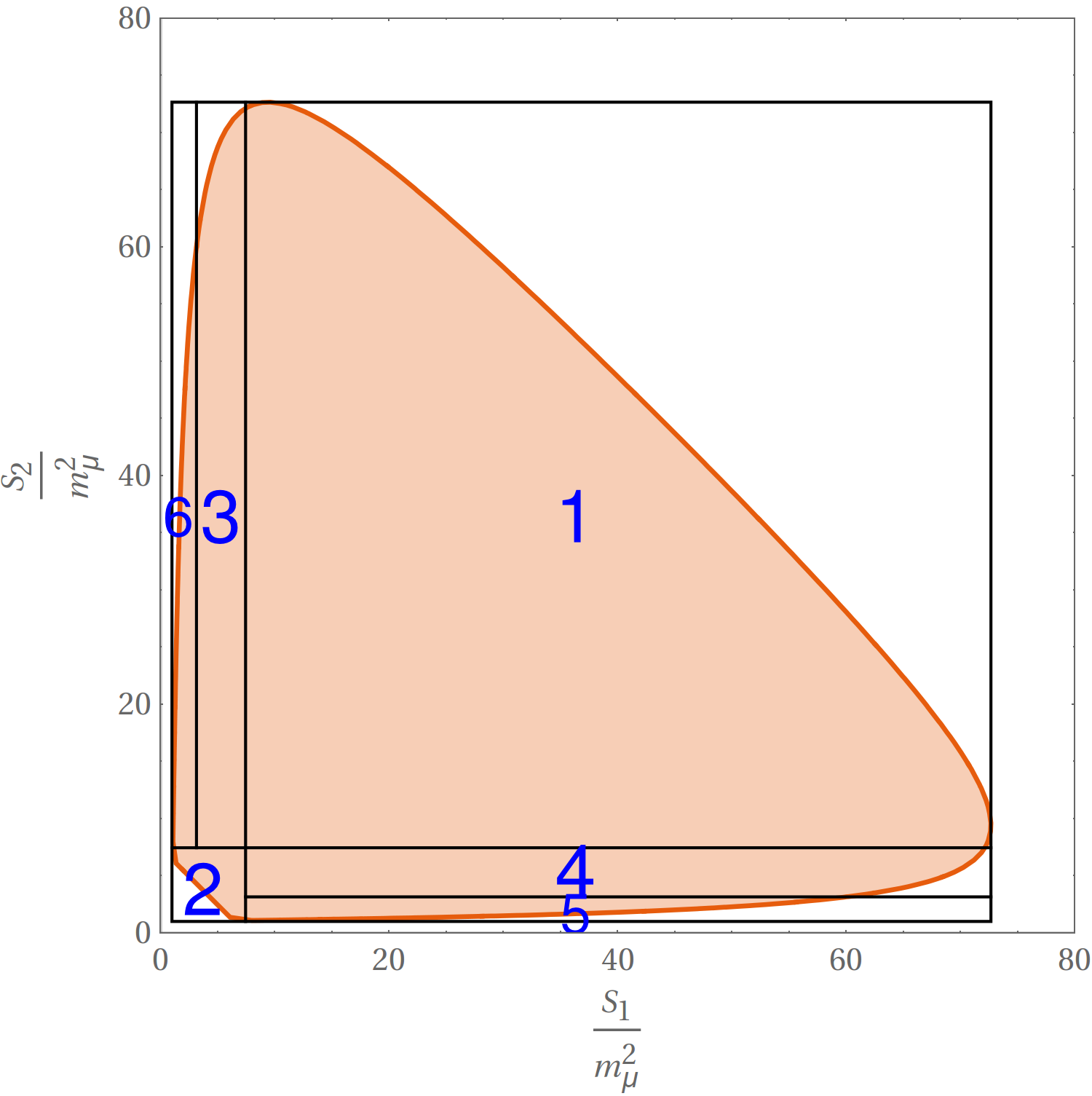}
	\caption{The figure shows the method of splitting into regions for integration in the $ s_1 $--$ s_2 $ plane for the energy  $ \sqrt{s}/2=500 $ MeV. Orange color indicates the physical region of the invariants $ s_1 $ and $ s_2 $. Regions No. 5, 6 give the main contribution to the NLO of the cross section.}\label{fig:reg}
\end{figure}

The numerical result was obtained using Monte-Carlo integration with the number of points $ N=50000 $ for Born and corrections $\delta_2,\,\delta_{3} $. In addition, the "warming up" of the Vegas algorithm with the number of points $ N=1000 $ was used. When calculating the correction to the cross section $ \delta_1 $, the integration region over the invariants $ s_1 $ and $ s_2 $ was divided into $ 6 $ regions (see.  figure~\ref{fig:reg}).  In each region the integration was carried out using the Monte Carlo method with the number of points  $ N=25000 $. The integration region has to be divided into $ 6 $ regions since the square of the amplitude is peak in the regions $ s_{1,2}\approx m_{\mu}^2 $. All numerical results were obtained using the supercomputer of the Novosibirsk State University NUSC~\cite{nusc}.

\begin{figure}[t]
	\includegraphics[width=8.2cm]{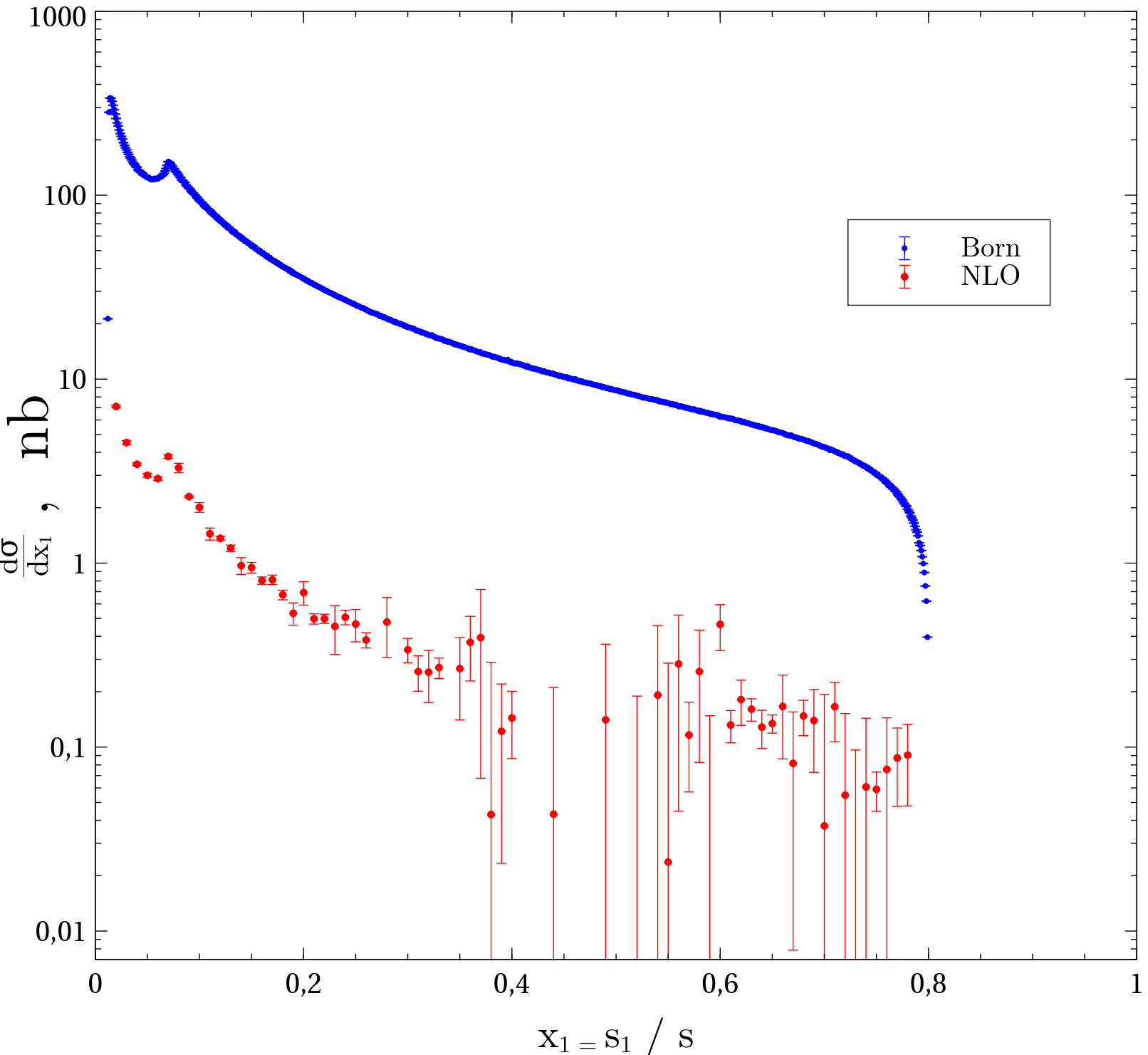}
	\includegraphics[width=7.6cm]{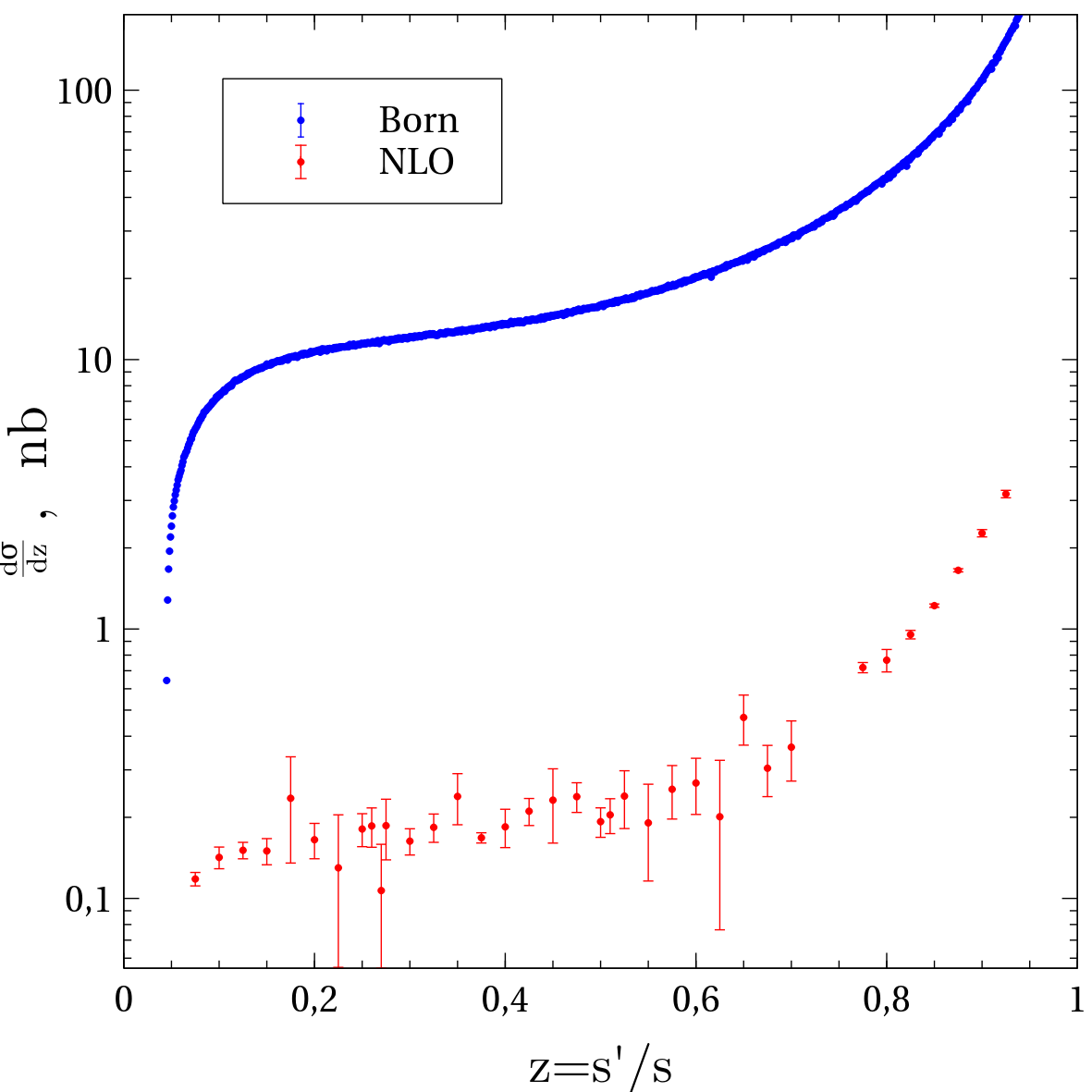}
	\caption{Differential cross sections vs the invariant mass of the muon and photon $ s_1=(p_1+k_1)^2 $ in the left picture. Differential cross sections vs the invariant mass of the muons $ s'=(p_1+p_2)^2 $ in the right picture.  The energy values in the figures are $ \sqrt{s}/2=500 $ MeV.}\label{fig:dist1}
\end{figure}
\begin{figure}[t]
	\includegraphics[width=8cm]{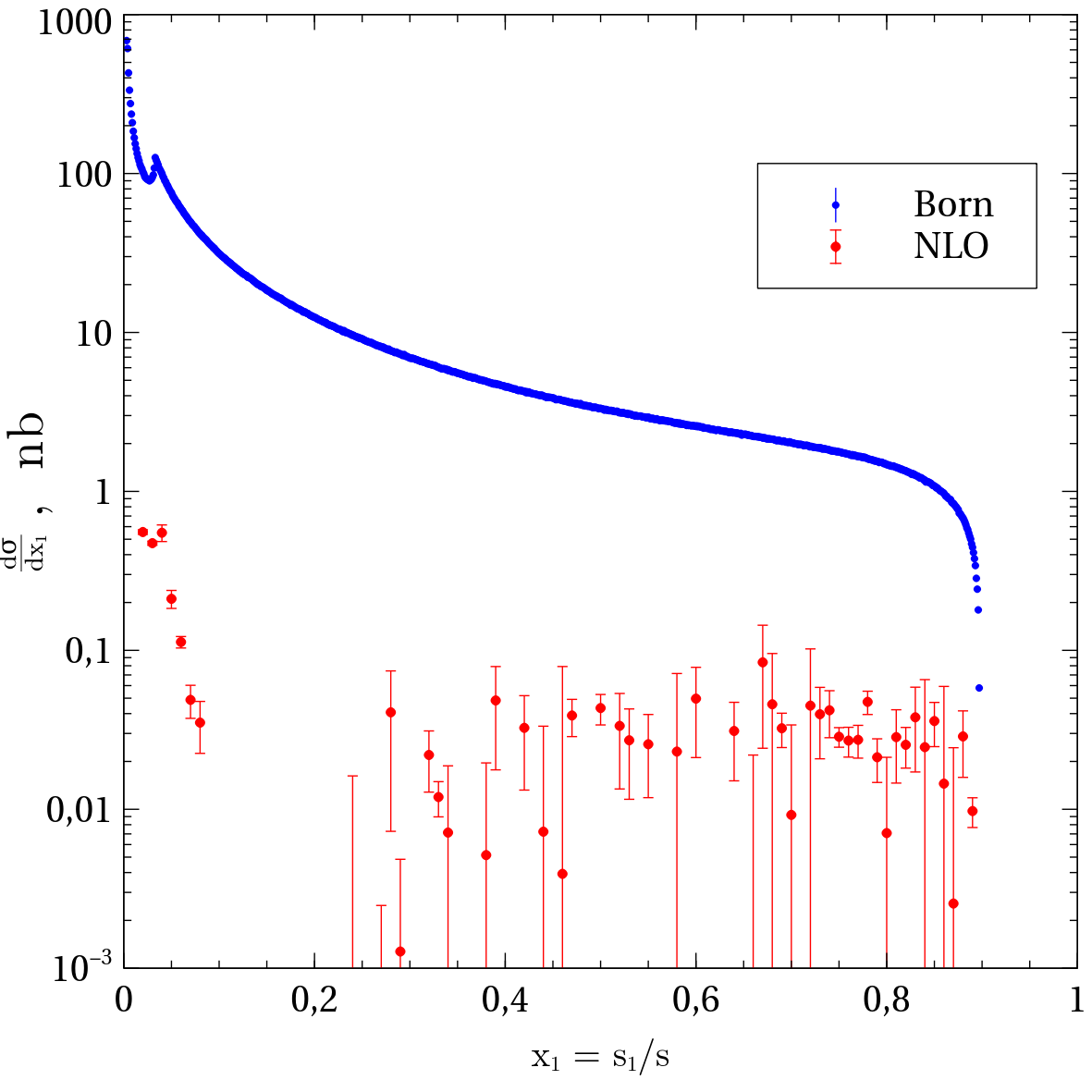}
	\includegraphics[width=8cm]{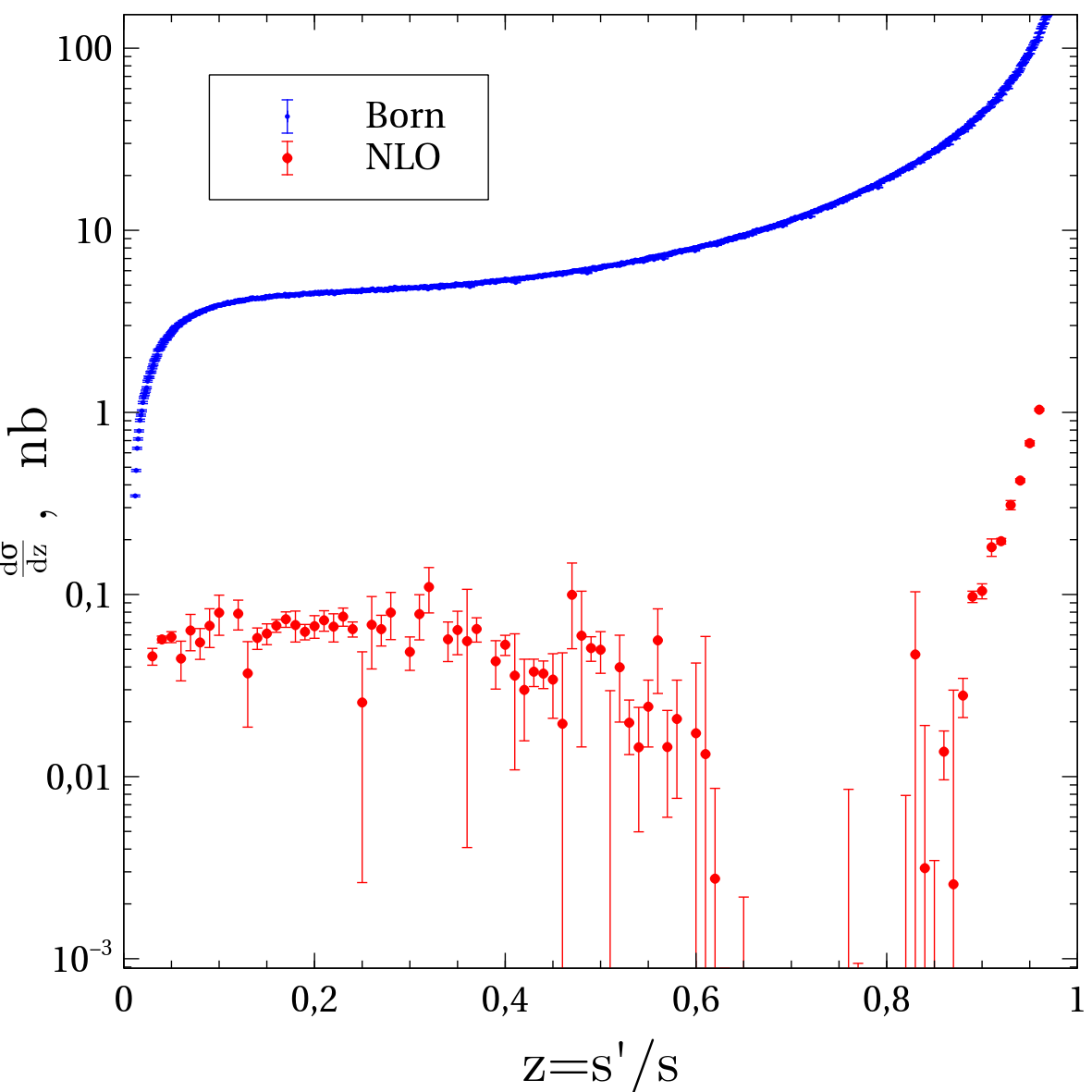}
	\includegraphics[width=8cm]{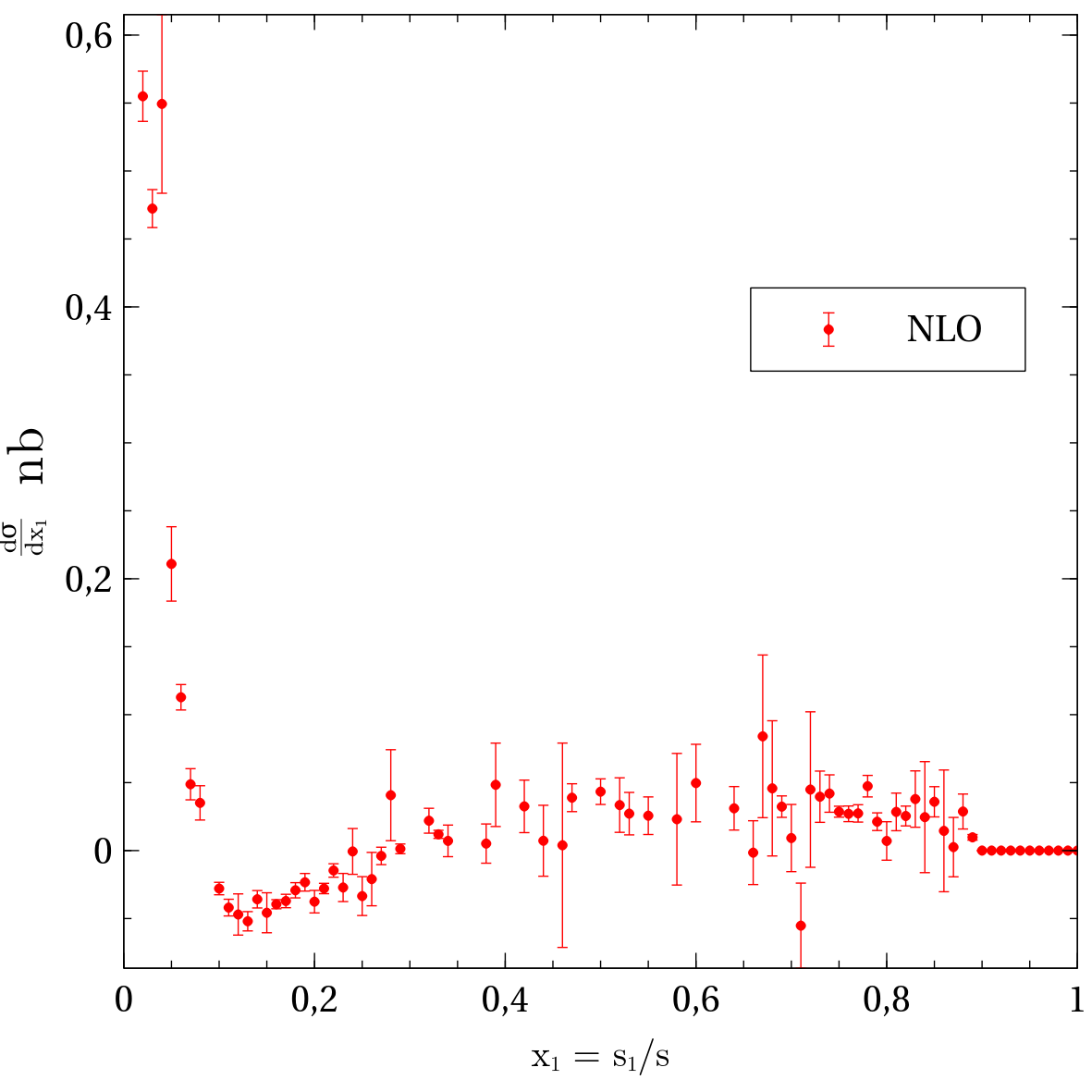}
	\includegraphics[width=8cm]{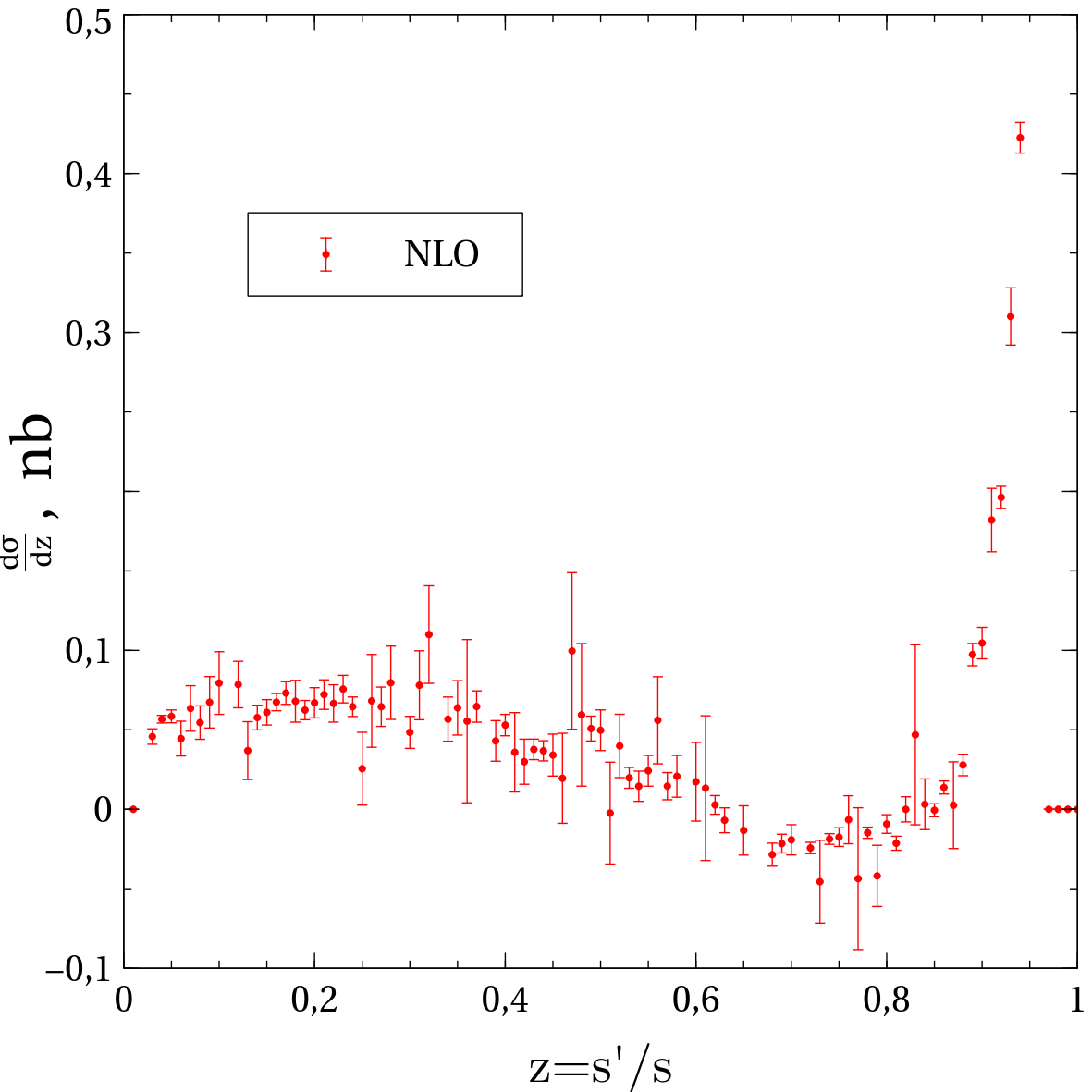}
	\caption{Differential cross sections vs the invariant mass of the muon and photon $ s_1=(p_1+k_1)^2 $ in the left pictures. Differential cross sections vs the invariant mass of the muons $ s'=(p_1+p_2)^2 $ in the right pictures.  The energy values in the figures are $ \sqrt{s}/2=1000 $ MeV. The upper pictures are on a logarithmic scale, and the lower ones are on a normal scale.}\label{fig:dist3}
\end{figure}

Differential cross sections were calculated for the invariant mass of the muon and photon $ s_1=(p_1+k)^2 $, as well as for the invariant mass of the two muons $ s'=(p_1+p_2)^2 $. The results can be seen in the figures~\ref{fig:dist1} and~\ref{fig:dist3}. Small peaks in the plots of the differential cross sections $ \frac{d\sigma}{dx_1} $ are associated with conditions for the minimum  energy of the final photon (see figure~\ref{fig:born}).

\section{Calculation of the next-to-leading order corrections}


The next-to-leading order corrections can be divided into three types~\eqref{eq:nlo}: virtual corrections with one fermion line, light-by-light corrections and real corrections (production of the additional soft photon).

\subsection{Correction with an additional bremsstrahlung photon }

Here we discuss the third term in~\eqref{eq:nlo}. This term was calculated partially by the soft-photon approximation method to extract the infrared divergence. We work in the dimensional regularization with dimension $ d=4-2\epsilon $. Therefore equation for $ T^{(3)}$ is
\begin{equation}
T^{(3)}(\boldsymbol{s},\epsilon,\omega_{\max},n)=T^{(3)}_{s.p.}(\boldsymbol{s},\epsilon,\omega_{*},n)+T^{(3)}_{born}(\boldsymbol{s},\omega_{*},\omega_{\max},n)
\end{equation}
where $ \omega_* $ is the energy of the bremsstrahlung photon, to which the soft-photon approximation is applied in the reference frame $ n^{\mu}=(1,0,0,0) $,
\begin{equation}
T^{(3)}_{s.p.}(\boldsymbol{s},\epsilon,\omega_{*},n)=\frac{2}{4}\sum_{pol}|M_{born}|^2\frac{1}{4\pi}\int_0^{\omega_{*}}\frac{d\omega}{\omega^{1+2\epsilon}}\Bigl(f_{12}(n)-f_{11}(n)-f_{22}(n)\Bigr)\,,
\end{equation}
\begin{equation}
T^{(3)}_{born}(\boldsymbol{s},\omega_{*},\omega_{\max},n)=\int\frac{d^3k'}{2\omega'(2\pi)^3}\frac{1}{4}\sum_{pol}|M_{\gamma\gamma\rightarrow \mu^+\mu^- \gamma\gamma}|^2\Theta\bigl(\omega_{\max}-(k'n)\bigr)\Theta\bigl((k'n)-\omega_*\bigr)\,.
\end{equation}
Here $ \omega_{\max} $ is the maximum energy of the bremsstrahlung photon in the reference frame $ n^{\mu}=(1,0,0,0) $ (four vector of the reference frame has normalization $ n^2=1 $) and $ f_{ij}(n) $ has the following form:
\begin{equation}
f_{ij}=\int\frac{\omega^{2\epsilon}d^{3-2\epsilon}q}{2|\vec{q}|(2\pi)^{1-2\epsilon}}\frac{2(p_i,p_j)}{(p_iq)(p_jq)}\delta\bigl((qn)/\omega-1\bigr)\,,
\end{equation}
where $ p_1,\,p_2 $ --- are momenta of the muons. After integration and expansion in $ \epsilon $ we have the following equation:
\begin{equation}
\begin{split}
&T_{s.p.}^{(3)}(\boldsymbol{s},\epsilon,\omega_*,n)=\frac{2}{4}\sum_{pol}|M_{born}|^2\frac{a_{\Gamma}}{\epsilon}\bigl(1-2\epsilon\ln( 2\omega_{*})\bigr)\Biggl((s-s_1-s_2)\frac{R_2(s-s_1-s_2+2)}{s-s_1-s_2-2}+\\
&+1-\epsilon\biggl(G\bigl(v,(np_1),x_1\bigr)-\frac{1}{2}F\bigl((np_1)\bigr)-\frac{1}{2}F\bigl((np_2)\bigr)\biggr)\Biggr)+{\cal O}(\epsilon)=\\
&=\frac{1}{\epsilon}\Phi(\boldsymbol{s},\epsilon)+T_+^{(3)}(\boldsymbol{s},\omega_*,n)+{\cal O}(\epsilon)\,,
\end{split}
\end{equation}
where we use the notations
\begin{equation}\label{eq:Phi}
\begin{split}
&a_{\Gamma}=\frac{\Gamma(1+\epsilon)}{(4\pi)^{2-\epsilon}}\frac{\Gamma^2(1-\epsilon)}{\Gamma(1-2\epsilon)}\,,\\
&v=\sqrt{1-\frac{4}{(s-s_1-s_2)^2}}\,,\quad x_1=\frac{1}{v}\Bigl((np_1)-\frac{2(np_2)}{s-s_1-s_2}\Bigr)\,,\\
&\Phi(\boldsymbol{s},\epsilon)=\frac{2}{4}a_{\Gamma}\sum_{pol}|M_{born}|^2\biggl((s-s_1-s_2)\frac{R_2(s-s_1-s_2+2)}{s-s_1-s_2-2}+1\biggr)\,,\\
&G(a,b,c)=\int_1^{-1}\frac{dx}{1-ax}\ln\frac{b-cx+\sqrt{1-x^2+(bx-c)^2}}{2\sqrt{1-x^2}}\,,\\
&F(x)=\frac{x}{\sqrt{x^2-1}}\ln\frac{1+\sqrt{1-\frac{1}{x^2}}}{1-\sqrt{1-\frac{1}{x^2}}}\,.
\end{split}
\end{equation}
The function $ R_2 $ is defined in Appendix~\eqref{eq:R2}.

\subsection{Corrections with one fermion line}
\begin{figure}[t]
	\includegraphics[width=15cm]{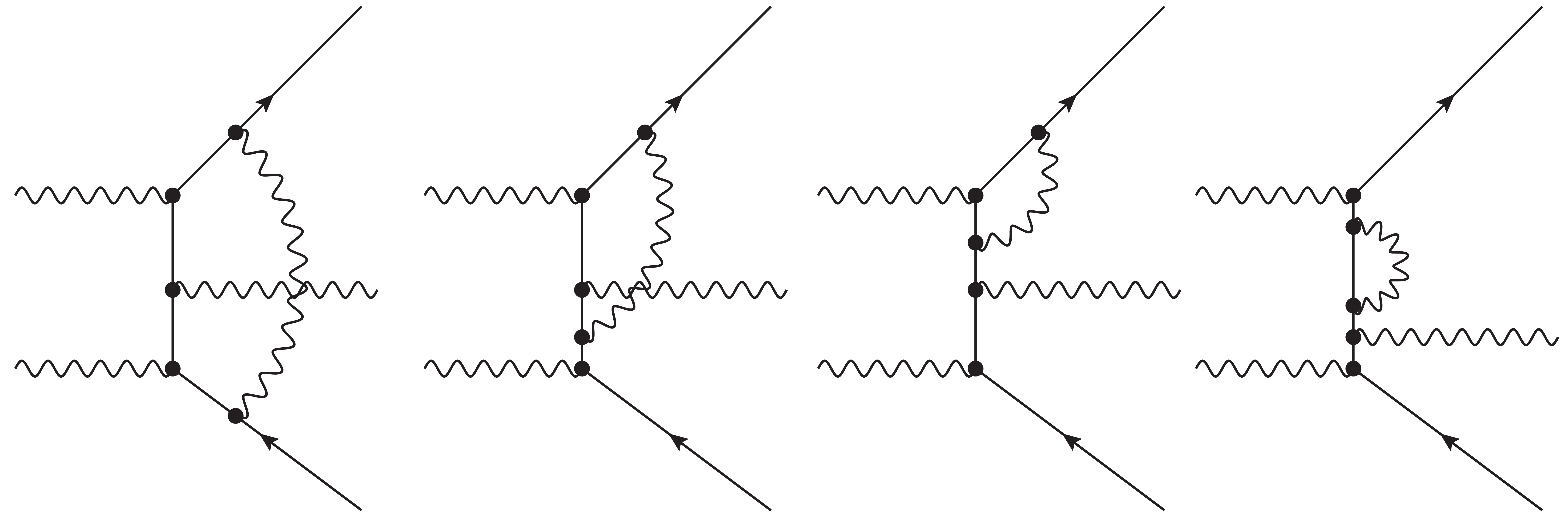}
	\caption{Examples of standard QED corrections diagrams with one fermion line}\label{fig:2}
\end{figure}

QED virtual corrections with one fermion line represent interference of the amplitude $ M_1 $ with the Born amplitude
\begin{equation}
T^{(1)}(\boldsymbol{s},\epsilon)=\frac{2}{4}\sum_{pol}\Re(M_1M_{born}^*)=-\frac{1}{\epsilon}\Phi(\boldsymbol{s},\epsilon)+T_+^{(1)}(\boldsymbol{s})+{\cal O}(\epsilon)\,.
\end{equation}
The amplitude $ M_1 $ is the sum of the Feynman diagrams with one fermion line (see fig.~\ref{fig:2}). The matrix element of amplitude $ M_1 $ contains infrared and ultraviolet divergences. The ultraviolet divergences are canceled by counterterms, and the infrared divergences are canceled by $ T^{(3)}(\boldsymbol{s},\epsilon,\omega_{\max},n) $ (see equations~\eqref{eq:ir_sing}).

Since all calculations are automated in the \textit{Wolfram Mathematica}, we calculated $ T^{(1)}(\boldsymbol{s},\epsilon) $ instead of $ M_1 $ to avoid tensor integrals. In $ M_1 $ there are $ 48 $ diagrams with one fermion line (for example see fig.~\ref{fig:2}). The algorithm for calculating $ T^{(1)}(\boldsymbol{s},\epsilon)  $ is as follows:
\begin{enumerate}
	\item We perform  summation over polarizations for all particles.
	\item We compute traces of gamma matrices.
	\item We reduce integrals with the loop momenta in the numerator to integrals without  numerator.
	\item We reduce loop integrals to master-integrals.
	\item Using the expressions for the basis integrals, we perform expansion in $ \epsilon $ where $ d=4-2\epsilon $ is dimension. 
	\item Finally, we cancel all divergent terms $ \sim 1/\epsilon $ and $\sim 1/\epsilon^2  $ using counterterms and corrections with the additional soft photon.	
\end{enumerate}
Algorithm steps 3 and 4 are performed using the package ``\textit{LiteRed}''~\cite{litered} for the \textit{Wolfram Mathematica}. In total, there are 6 families of master-integrals due to the permutational symmetry of photons, and in each family of integrals there are 21 master-integrals see fig.~\ref{fig:pmasters}.

\subsection{Light by light QED corrections}
\begin{figure}[t]
	\includegraphics[width=8cm]{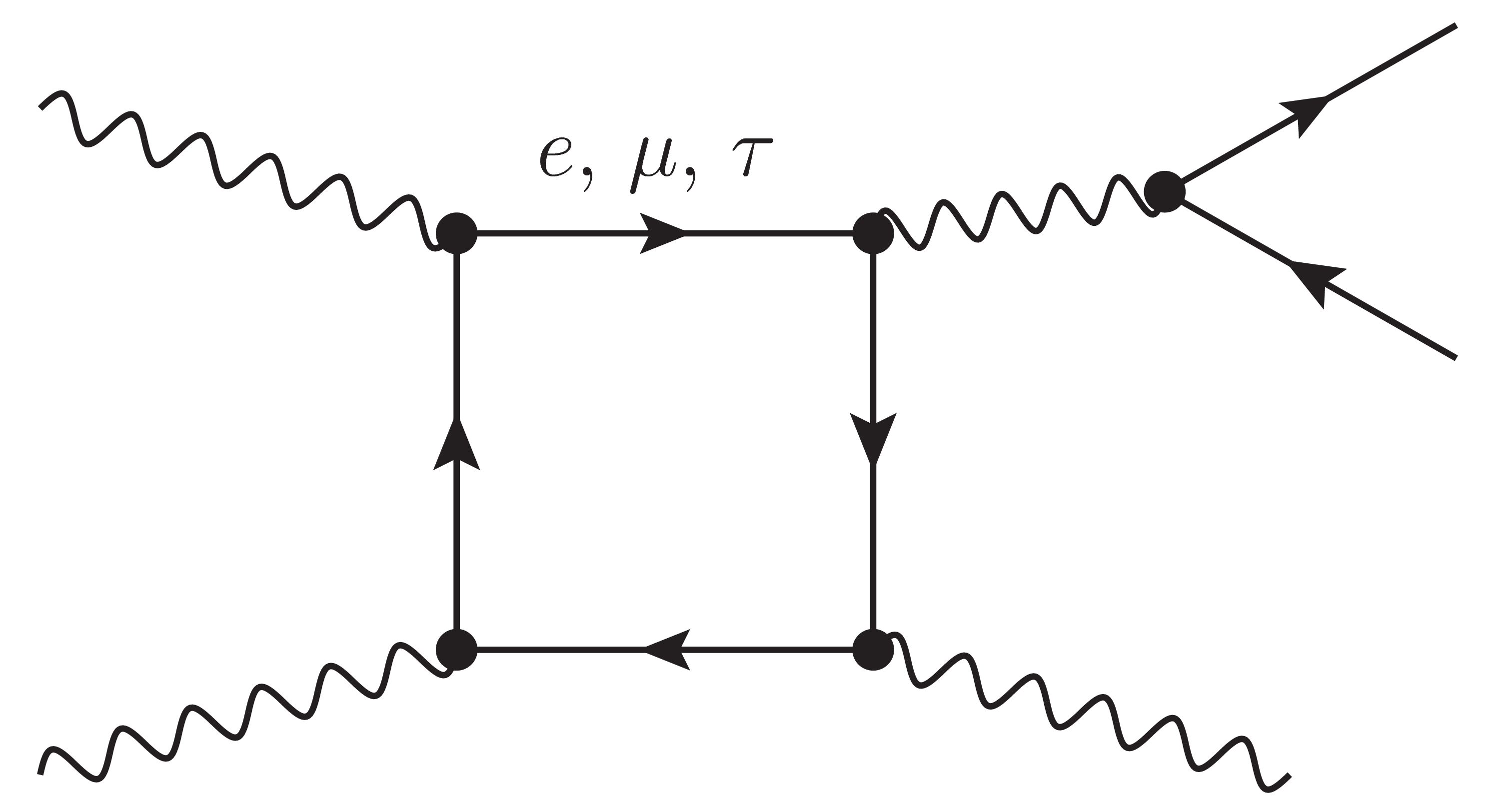}
	\caption{Example of QED light by light scattering diagram}\label{fig:lbl}
\end{figure}

QED corrections with two fermion lines are expressed in terms of the following sum
\begin{equation}
T^{(2)}(\boldsymbol{s})=\sum_{f=e,\mu,...}T^{(2)}(\boldsymbol{s},m_f).
\end{equation}
Here $ m_i $ is the fermion mass and 
\begin{equation}
T^{(2)}(\boldsymbol{s},m_f)=\frac{2}{4}\Re\sum_{pol} M_2(\boldsymbol{s},m_f)M_{born}^*\,,
\end{equation}
\begin{equation}\label{eq:M2}
M_2(\boldsymbol{s},m_f)=-Q_f^4e^5\bar{u}(p_1)\gamma^{\mu}v(p_2)e^{*\nu}(k)e^{\alpha}(k_1)e^{\beta}(k_2)V_{\mu\nu\alpha\beta}
\end{equation}
where $ M_2 $ is the amplitude with two fermion lines, for example see fig.~\ref{fig:lbl}. There are six diagrams in $ M_2 $. They are expressed through three diagrams, which can be obtained from figure~\ref{fig:lbl} by permuting photons. The amplitude $ M_2 $ is finite in dimension $ d=4-2\epsilon $, but individual diagrams contain ultraviolet divergences. Unlike the corrections with one fermion line, these corrections contain tensor integrals. All the tensor integrals are expressed in terms of integrals with numerators that depend on the scalar products of the loop momenta and the momentum in the denominator~\eqref{eq:tensor}.

For this type of corrections, we first calculate the tensor $ V_{\mu\nu\alpha\beta} $. The algorithm for calculation of the tensor $ V_{\mu\nu\alpha\beta} $ is
\begin{enumerate}
	\item We compute traces of gamma matrices.
	\item We reduce integrals with the loop momentum in the numerator to integrals without numerator.
	\item We reduce the tensor integrals to integrals with a numerator reducible to the denominator~\eqref{eq:tensor} and again perform step number 2.
	\item We reduce all loop integrals to basis master integrals.
	\item Using the expressions for the basis integrals, we perform the expansion in $ \epsilon $.
	\item Finally we cancel the ultraviolet divergences between different diagrams. 
\end{enumerate}
Loop integrals are reduced to three families of basis master integrals, see Appendix~\eqref{eq:mcb1}. Further the algorithm for calculation of  $ T^{(2)}(\boldsymbol{s},m_i) $ is
\begin{enumerate}
	\item We perform summation over polarization  in the following expression\\ $\sum_{pol}\bar{u}(p_1)\gamma^{\mu}v(p_2)e^{*\nu}(k)e^{\alpha}(k_1)e^{\beta}(k_2)M^*_{born} $.
	\item Finally we perform convolution of indices in~\eqref{eq:M2}.
\end{enumerate}

\section{Calculation of the cross section}

The expression for the scattering cross section can be represented as follows~\cite{BK:1973}:
\begin{equation}\label{eq:sigma}
\begin{split}
&\sigma=\frac{1}{2s}\int\Bigl(\frac{1}{4}\sum_{pol}|M|^2\Bigr)d\Gamma_3=\\
&=\frac{1}{2s(2\pi)^5}\int\delta^{(4)}(k_1+k_2-p_1-p_2-k)\Bigl(\frac{1}{4}\sum_{pol}|M|^2\Bigr)\frac{d^3p_1}{2\epsilon_1}\frac{d^3p_2}{2\epsilon_2}\frac{d^3k}{2\omega}=\\
&=\frac{1}{2s(2\pi)^5}\int \frac{\pi\, dt_1ds_2}{8s\sqrt{\lambda(s,s_2,m^2)}}\Theta\bigl[-G(s,t_1,s_2,0,0,m^2)\bigr]\times\\
&\qquad\times\int ds_1\Theta\bigl[-G(s_1,s_2,s,0,m^2,m^2)\bigr]\int_0^{2\pi}d\lambda_1\Bigl(\frac{1}{4}\sum_{pol}|M|^2\Bigr)
\end{split}
\end{equation}
where $ \Theta(x) $ is the Heaviside function, $ \lambda_1 $ is the spiral angle:
\begin{equation}
\cos\lambda_1=\frac{\bigl([\vec{k}_1\times\vec{p}_1],[\vec{p}_1\times\vec{p}_2]\bigr)}{|[\vec{k}_1\times\vec{p}_1]|\;|[\vec{p}_1\times\vec{p}_2]|}
\end{equation}
in the center of mass frame of the initial photons ($ \vec{k}_1+\vec{k}_2=0 $), 
\begin{equation}
\begin{split}
&G(x,y,z,u,v,w)=v^2w + u^2z + u\bigl((w - x)(y-v) - (v + w + x + y)z + z^2\bigr)+\\
& + x\bigl(y(x + y - z) + w(z-y)\bigr) + v\bigl(w^2 + y(z-x) - w(x + y + z)\bigr)\,.
\end{split}
\end{equation}
This function is related to the Gram determinant
\begin{equation}
\Delta_3(p_1,p_2,p_3)=-\frac{1}{4}G\bigl((p_1+p_2)^2,(p_1-p_3)^2,(p_1+p_2-p_3)^2,p_1^2,p_2^2,p_3^2\bigr) \,.
\end{equation}
The function $ \lambda(x,y,z) $ is
\begin{equation}
\lambda(x,y,z)=(x-y-z)^2-4yz\,.
\end{equation}
The invariant $ t_2 $  is expressed through invariants and the spiral angle $ \lambda_1 $ as follows:
\begin{equation}
\begin{split}
&t_2=1+\frac{2\sqrt{G(s,t_1,s_2,0,0,m^2)G(s_1,s_2,s,0,m^2,m^2)}\cos(\lambda_1)}{\lambda(s,s_2,m^2)} +\\
&+\frac{ (s_1-m^2)(s_2-m^2)(s_2-t_1) + s^2(t_1-m^2) - s((2m^2 + s_1)t_1 + s_2(s_1 + t_1-4m^2)-m^4) }{\lambda(s,s_2,m^2)}\,.
\end{split}
\end{equation}
The representation of the scattering cross section in equation~\eqref{eq:sigma} was used for numerical calculation. The integration was carried out by the Monte Carlo method using Vegas algorithm~\cite{vegas} (implemented in \verb|GSL| library) over 4-dimensional space.

\section{Conclusion}
The NLO QED corrections to the differential cross section of the process $ \gamma\gamma\rightarrow\mu^+\mu^-\gamma $ have been calculated. The result is obtained in the analytical form and implemented as functions in the C programming language~\cite{ggmmg}. Numerical results were also obtained for corrections to total and differential cross sections. The relative value of the correction to the cross section is of the order of $\sim 1-2\% $.

\section*{Acknowledgments}
This work is supported in part by the Russian Foundation for
Basic Research Grants No. 16-32-60033 and 15-02-09016. Numerical results were obtained using a supercomputer of the Novosibirsk State University NUSC~\cite{nusc}.

\section*{Appendix}

\subsection*{Description of functions}

Here we discuss the implementation of functions that are the result of calculations. Source code can be taken from~\cite{ggmmg}. The function 
\begin{equation}
T_{2\mu\gamma}^{(born)}(\boldsymbol{s})=\sum_{pol}|M_{born}|^2
\end{equation}
describes the main contribution to the differential cross section of the process $ \gamma\gamma\rightarrow\mu^+\mu^-\gamma $. It is implemented as \verb|double born(double *vars)|, where \verb|double *vars| is an array of five numbers: 
\[
\begin{array}{ccccc}
\verb|vars[0]| & \verb|vars[1]| & \verb|vars[2]|& \verb|vars[3]|& \verb|vars[4]| \\ 
\dfrac{s}{m_{\mu}^2} & \dfrac{s_1}{m_{\mu}^2} &  \dfrac{s_2}{m_{\mu}^2} & \dfrac{t_1}{m_{\mu}^2} & \dfrac{t_2}{m_{\mu}^2}\,. \\
\end{array}
\]

The function $ T_+^{(1)}(\boldsymbol{s}) $ is implemented as two different functions:
\[
T^{(1)}_+(\boldsymbol{s})=T^{(1.1)}(\boldsymbol{s})+T^{(1.2)}(\boldsymbol{s})
\]
where functions $ T^{(1.1)},\; T^{(1.2)}$ describe corrections with a single fermion line. Several box diagrams with a single fermion line are implemented in the function $  T^{(1.2)} $, since integrated over the invariants of the finite particles, they give a large error. Therefore they must be integrated separately. Functions $ T^{(1.1)},\; T^{(1.2)}$ are implemented as
\verb|double nlo1(double *vars)| and \verb|double nlo1_2(double *vars)| 
where \verb|double *vars| is an array of five numbers: 
\[
\begin{array}{ccccc}
 \verb|vars[0]| & \verb|vars[1]| & \verb|vars[2]|& \verb|vars[3]|& \verb|vars[4]| \\ 
\dfrac{s}{m_{\mu}^2} & \dfrac{s_1}{m_{\mu}^2} &  \dfrac{s_2}{m_{\mu}^2} & \dfrac{t_1}{m_{\mu}^2} & \dfrac{t_2}{m_{\mu}^2} \\
\end{array}
\]
The function $T_+^{(3)}(\boldsymbol{s},\omega_*,n)$ in the reference frame $ n^{\mu}=(1,0,0,0) $ is implemented as\\
\verb|double nlo3(double *vars)| were  \verb|double *vars| is an array of six numbers: 
\[
\begin{array}{cccccc}
\verb|vars[0]| & \verb|vars[1]| & \verb|vars[2]|& \verb|vars[3]|& \verb|vars[4]|& \verb|vars[5]| \\ 
\dfrac{s}{m_{\mu}^2} & \dfrac{s_1}{m_{\mu}^2} &  \dfrac{s_2}{m_{\mu}^2} & \dfrac{t_1}{m_{\mu}^2} & \dfrac{t_2}{m_{\mu}^2} & \dfrac{\omega_*}{m_{\mu}}\\
\end{array}
\]
Here $ \omega_* $ is the maximum energy of the soft photon for soft photon approximation in the $ n^{\mu}=(1,0,0,0)$ reference frame. For an arbitrary reference frame the function $ 2\times T_+^{(3)}(\boldsymbol{s},\omega_*,n) $ is implemented as\\
\verb|double ir(double s,double s1,double s2,double t1,double t2,| \\
\verb|          double E,double np1, double np2)|\\
where
\[
\begin{array}{cccccccc}
\verb|s| & \verb|s1| & \verb|s2|& \verb|t1|& \verb|t2| &\verb|E|&\verb|np1| &\verb|np2| \\ 
\dfrac{s}{m_{\mu}^2} & \dfrac{s_1}{m_{\mu}^2} &  \dfrac{s_2}{m_{\mu}^2} & \dfrac{t_1}{m_{\mu}^2} & \dfrac{t_2}{m_{\mu}^2} & \dfrac{\omega_*}{m_{\mu}} & \dfrac{(n,p_1)}{m_{\mu}}&\dfrac{(n,p_2)}{m_{\mu}} \\
\end{array}
\]
$ (n,p_1),\,(n,p_2) $ are scalar products of the four-vectors of muons $ p_1,\,p_2 $ with the four vector of the reference frame $ n^{\mu} $.

The function $ T^{(2)}(\boldsymbol{s},m) $ is implemented as \verb|double nlo2(double *vars)|, where \verb|double *vars| is an array of six numbers: 
\[
\begin{array}{cccccccc}
\verb|vars[0]| & \verb|vars[1]| & \verb|vars[2]|& \verb|vars[3]|& \verb|vars[4]| &\verb|vars[5]|\\ 
\dfrac{s}{m_{\mu}^2} & \dfrac{s_1}{m_{\mu}^2} &  \dfrac{s_2}{m_{\mu}^2} & \dfrac{t_1}{m_{\mu}^2} & \dfrac{t_2}{m_{\mu}^2} & \dfrac{m}{m_{\mu}} \\
\end{array}
\]
Here $ m $ is the mass of virtual fermion particle. \textbf{The value returned by all  functions must be multiplied by $ (4\pi)^4\dfrac{\alpha^4}{8m_{\mu}^2} $ for the final result}.

\subsection*{Master-integrals}

\subsubsection*{Master-integrals for corrections with one fermion line}

\begin{figure}[t]
		\includegraphics[width=17cm]{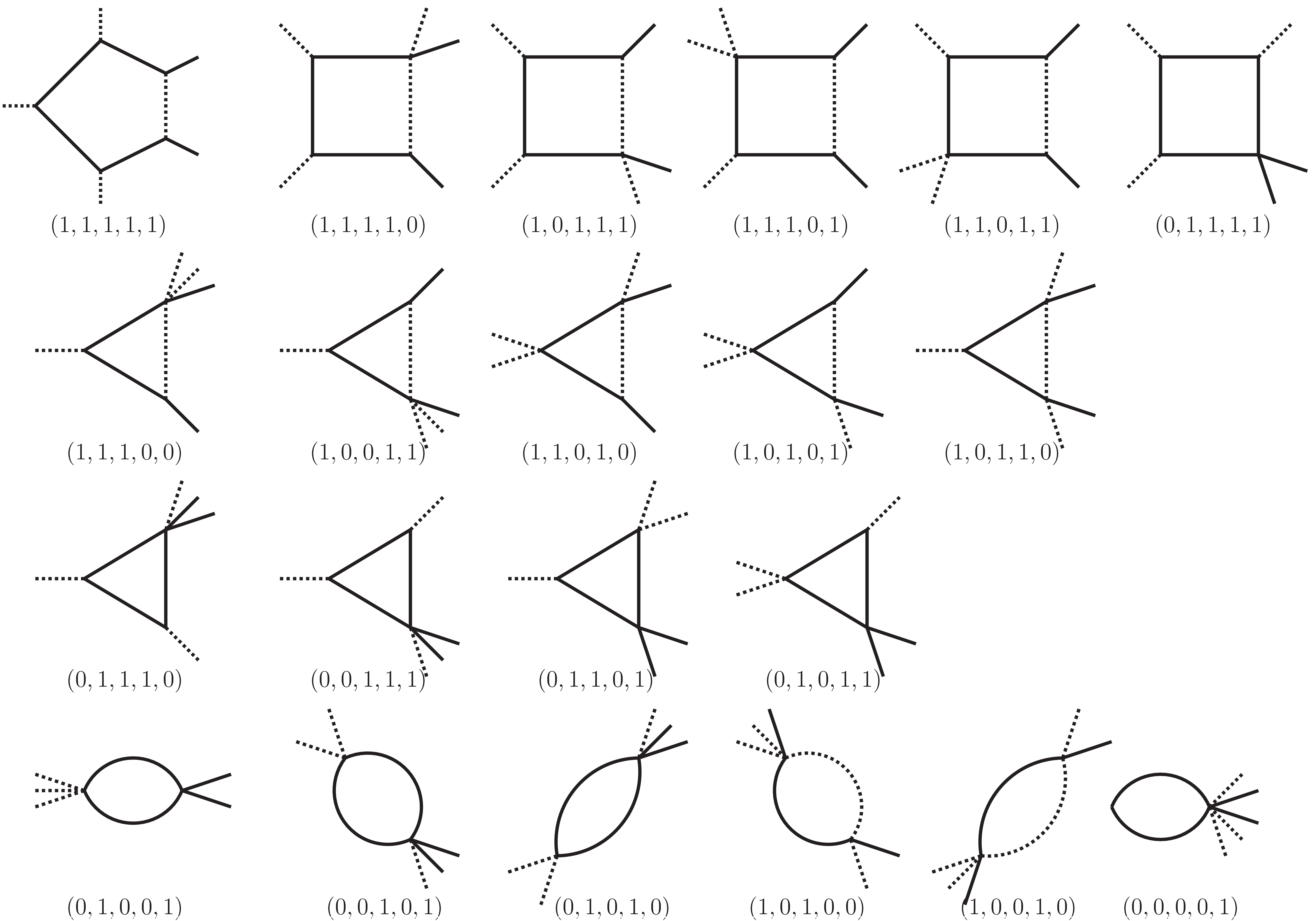}
	\caption{Basis of master-integrals. The signature under the diagram is the vector $ \vec{n} $. Doted line is the photon line.}\label{fig:pmasters}
\end{figure}

For corrections with one fermion line there are six families of master integrals,  which are distinguished by a permutation of the photon momenta. The first family of master integrals has the following topology:
\begin{equation}
\begin{split}
&J_{123}(n_1,n_2,n_3,n_4,n_5)=J_{123}(\vec{n})=\int\frac{d^{D}l}{(2\pi)^D}\frac{1}{D_1^{n_1}D_2^{n_1}D_3^{n_1}D_4^{n_1}D_5^{n_1}}\\
&D_1=l^2\,,\;D_2=(l-p_2)^2-1\,,\;D_3=(l-p_2+k_1)^2-1\,,\\
&D_4=(l-p_2+k_1+k_2)^2-1\,,\;D_5=(l-p_2+k_1+k_2-k)^2-1\,.
\end{split}
\end{equation}
There are twenty one basis master-integrals in this topology (see fig.~\ref{fig:pmasters}).
The remaining topologies are obtained by the following changes of the photon momenta in $ J_{123} $:
\begin{equation}
\begin{array}{cc}
\text{topology} & \text{changes}\\
\hline
J_{213} & k_1\leftrightarrow k_2\\
J_{132}	& k_2\leftrightarrow (-k)\\
J_{321} & k_1\leftrightarrow (-k)\\
J_{231} & k_1 \rightarrow k_2,\;k_2\rightarrow (-k),\; (-k)\rightarrow k_1\\
J_{312} & k_1 \rightarrow (-k),\; k_2\rightarrow k_1,\; (-k)\rightarrow k_2
\end{array}
\end{equation}
There are many basis master-integrals, but they are expressed in terms of a smaller number of functions. Master integrals with one and two denominators are
\begin{equation}
\begin{split}
&J_{123}(0,0,0,0,1)=a_{\Gamma} \frac{1+\epsilon}{\epsilon}+{\cal O}(\epsilon),\\
&J_{123}(1,0,1,0,0)=a_{\Gamma} \biggl(\frac{1}{\epsilon}+2+R_1(s_1-s-t_2+1)\biggr)+{\cal O}(\epsilon),\\
&J_{123}(1,0,0,1,0)=a_{\Gamma} \biggl(\frac{1}{\epsilon}+2+R_1(s_1)\biggr)+{\cal O}(\epsilon),\\
&J_{123}(0,1,0,1,0)=a_{\Gamma} \biggl(\frac{1}{\epsilon}+2+R_2(s)\biggr)+{\cal O}(\epsilon),\\
&J_{123}(0,1,0,0,1)=a_{\Gamma} \biggl(\frac{1}{\epsilon}+2+R_2(s-s_1-s_2+2)\biggr)+{\cal O}(\epsilon),\\
&J_{123}(0,0,1,0,1)=a_{\Gamma} \biggl(\frac{1}{\epsilon}+2+R_2(1-s_2+t_1-t_2)\biggr)+{\cal O}(\epsilon).
\end{split}
\end{equation}
Master-integrals with three denominators are
\begin{equation}
\begin{split}
&J_{123}(1,1,1,0,0)=a_{\Gamma} T_4(-s+s_1-t_2)+{\cal O}(\epsilon),\\
&J_{123}(1,1,0,1,0)=a_{\Gamma} T_2(s_1,s)+{\cal O}(\epsilon),\\
&J_{123}(1,0,1,1,0)=a_{\Gamma} T_3(1-(s-s_1+t_2),s_1)+{\cal O}(\epsilon),\\
&J_{123}(1,0,1,0,1)=a_{\Gamma} T_2(1-(s-s_1+t_2),1-s_2+t_1-t_2)+{\cal O}(\epsilon),\\
&J_{123}(1,0,0,1,1)=a_{\Gamma} T_4(s_1)+{\cal O}(\epsilon),\\
&J_{123}(0,1,1,1,0)=a_{\Gamma} T_5(s)+{\cal O}(\epsilon),\\
&J_{123}(0,1,1,0,1)=a_{\Gamma} T_6(s-s_1-s_2+2,1-s_2+t_1-t_2)+{\cal O}(\epsilon),\\
&J_{123}(0,1,0,1,1)=a_{\Gamma} T_6(s,s-s_1-s_2+2)+{\cal O}(\epsilon),\\
&J_{123}(0,0,1,1,1)=a_{\Gamma} T_5(1-s_2+t_1-t_2)+{\cal O}(\epsilon).
\end{split}
\end{equation}
Master-integrals with four denominators (boxes) are
\begin{equation}
\begin{split}
&J_{123}(1,1,1,1,0)=a_{\Gamma} B_1(1-(s-s_1+t_2),s,s_1)+{\cal O}(\epsilon),\\
&J_{123}(1,1,1,0,1)=a_{\Gamma} \biggl( \frac{R_2(s-s_1-s_2+2)}{\epsilon(-s+s_1-t_2)(s-s_1-s_2-2)}+\\
&\mspace{100mu}+B_2(s-s_1-s_2+2,-s+s_1-t_2+1,1-s_2+t_1-t_2)\biggr)+{\cal O}(\epsilon),\\
&J_{123}(1,1,0,1,1)=a_{\Gamma} \biggl( \frac{R_2(s-s_1-s_2+2)}{\epsilon(s_1-1)(s-s_1-s_2-2)}+\\
&\mspace{200mu}+B_2(s-s_1-s_2+2,s_1,s)\biggr)+{\cal O}(\epsilon),\\
&J_{123}(1,0,1,1,1)=a_{\Gamma} B_1(s_1,-(s_2-t_1+t_2-1),1-(s-s_1+t_2))+{\cal O}(\epsilon),\\
&J_{123}(0,1,1,1,1)=a_{\Gamma} B_3(s,-(s_2-t_1+t_2-1),s-s_1-s_2+2)+{\cal O}(\epsilon).
\end{split}
\end{equation}
We expressed twenty-one master integrals through ten functions. Since we need only the real part of the corrections to the scattering cross section, all the functions that will be given below contain only the real part of the master integrals. Functions $ R_i,\,T_i,\,B_i $ are
\begin{equation}\label{eq:R1}
R_1(s)=\Bigl(\frac{1}{s}-1\Bigr)\ln|1-s|\,,
\end{equation}
\begin{equation}\label{eq:R2}
R_2(s)=
\begin{cases}
-\sqrt{1-4/s}\ln\frac{1+\sqrt{1-4/s}}{1-\sqrt{1-4/s}}, & s\notin [0,4]\\
\sqrt{4/s-1}(2\arctan\sqrt{4/s-1}-\pi) &, s\in [0,4]
\end{cases}\,,
\end{equation}
\begin{equation}
T_3(t_1,t_2)=-\frac{\Li_2(t_1)-\Li_2(t_2)}{t_1-t_2}\,,\;
T_4(s)=\frac{1}{s-1}\Bigl(\frac{\pi^2}{6}-\Li_2(s)\Bigr),
\end{equation}
\begin{equation}
T_5(s)=\frac{\lnb(s)}{2s}\,,\;T_6(s,t)=\frac{\lnb(s)-\lnb(t)}{2(s-t)}\,,
\end{equation}
where we introduce new function:
\begin{equation}
\lnb(x)\equiv
\begin{cases}
\ln^2\frac{1+\sqrt{1-4/x}}{\sqrt{1-4/x}-1}, & x<0\\
\ln^2\frac{1+\sqrt{1-4/x}}{1-\sqrt{1-4/x}}-\pi^2, & x>4\\
-\Bigl(\pi-2\arctan(\sqrt{4/x-1}\Bigr)^2, & x\in [0,4]
\end{cases}\;.
\end{equation}
The most complicated function for master integral with three propagators reads:
\begin{equation}
\begin{split}
&T_2(t_1,t_2)=\frac{1}{D_3}\biggl(S_3(x_1,1,1/t_1)-S_3(x_2,1,1)+S_3(x_3,x_{31},x_{32})\biggr),\\
&D_3=\sqrt{(t_1+t_2-1)^2-4t_1t_2},\quad x_1=\frac{1-t_1^2-t_2+t_1t_2+(1+t_1)D_3}{2t_1D_3},\\
&x_{2,3}=-\frac{-2+2t_1+3t_2+t_1t_2-t_2^2+(t_2-2)D_3}{D_3(1\pm t_1-t_2+D_3)}\,,\;x_{31,32}=\frac{t_2\pm \sqrt{(t_2-4)t_2}}{2t_2}\,,\\
&T_2(1,t_2)=\frac{1}{D_3}S_3(x_3,x_{31},x_{32})
\end{split}
\end{equation}
here we use function
\begin{equation}\label{eq:S3}
S_3(x,a,b)=R(x,a)+R(x,b)\,,\quad
R(x,y)=\Li_2\Bigl(\frac{x}{x-y}\Bigr)-\Li_2\Bigl(\frac{x-1}{x-y}\Bigr)\,.
\end{equation}
Next we introduce functions used in the box master integrals:
\begin{equation}
\begin{split}
&B_1(s,t,m)=\int_0^1\frac{dy}{y(1-y)s(t-1)+m-t}\ln\biggl(\frac{t-1}{m-1}\bigl(1-y(1-y)s\bigr)\biggr)=\\
&=\frac{1}{s(1-t)\sqrt{D}}\biggl(S_3(y_{1+},y_{2+},y_{2-})-S_3(y_{1-},y_{2+},y_{2-})\biggr),\\
&D=1+\frac{4(m-t)}{s(t-1)}\,,\quad y_{1\pm}=\frac{1}{2}\bigl(1\pm\sqrt{D}\bigr)\,,\quad y_{2\pm}=\frac{1}{2}\Bigl(1\pm\sqrt{1-\frac{4}{s}}\Bigr)\,,
\end{split}
\end{equation}
\begin{equation}
\begin{split}
&B_2(s,t,m)=\frac{2}{s(t-1)\sqrt{1-4/s}}\biggl(\frac{\pi^2}{6}-2\Bigl(\Phi\bigl(X(s),1/X(m)\bigr) + \Phi\bigl(X(s),X(m)\bigr)\Bigr)+\\
& + \pi^2\Theta\bigl(-X(m)\bigr)-\Li_2\bigl(X^2(s)\bigr)-\ln^2\bigl(X(m)\bigr)-\\
&- 2\Bigl(-\pi^2\Theta(t-1)\Theta\bigl(-X(s)\bigr) + \ln(1-t)\ln\bigl(X(s)\bigr)\Bigr) - 2\ln\bigl(X(s)\bigr)\ln\bigl(1-X^2(s)\bigr)\biggr),  \\
&X(s)=\frac{\sqrt{1-4/s}-1}{\sqrt{1-4/s}+1},\\
&\Phi(x,y)=\pi^2\bigl(\Theta(-x) + \Theta(-y)\bigr)\Theta(xy-1) + \Theta(-xy)\Bigl(\frac{\pi^2}{6}-\Li_2(xy)\Bigr)-\\
&-\bigl(\ln(x) + \ln(y)\bigr)\ln(1-xy)+\Theta(xy)\bigl(\Li_2(1-xy)+\ln(xy)\ln(1- xy)\bigr),
\end{split}
\end{equation}
\begin{equation}\label{eq:B3}
\begin{split}
&B_3(s,t,m)=\frac{1}{s t \sqrt{1+4 (m-s-t)/(s t)}}\sum_{\sigma=\pm}\sigma\Bigl\{R(y_{2\sigma},0)- S_3(y_{2\sigma},x_{3+},x_{3-})-\\
&\qquad-R\Bigl(y_{2\sigma},\frac{t}{t+s-m}\Bigr)-R(1+y_{1\sigma},0)+S_3(1+y_{1\sigma},x_{1+},x_{1-})-R(-y_{1\sigma},1)+\\
&\qquad+S_3(-y_{1\sigma},x_{2+},x_{2-})\Bigr\},\\
&y_{1\pm}=\frac{1}{2}\Bigl(-1\pm \sqrt{1+4 (m-s-t)/(s t)}\Bigr)\,,\;
y_{2\pm}=\frac{t\bigl(1\pm\sqrt{1+4 (m-s-t)/(s t)}\bigr)}{2(s+t-m)}\,,\\
&x_{1\pm}=\frac{1}{2}(1\pm\sqrt{1-4/m})\,,\;
x_{2\pm}=\frac{1}{2}(1\pm\sqrt{1-4/t})\,,\;
x_{3\pm}=\frac{1}{2}(1\pm\sqrt{1-4/s}).
\end{split}
\end{equation}
Expressions for loop integrals can be found in~\cite{qcdloop}, except $ T_2 $, $ B_1 $ and $ B_3 $. Expressions for $ T_2,\,B_1,\,B_3 $ can be obtained using~\cite{TV:1979}.

The most difficult master integral is expressed through a linear combination of boxes
\begin{equation}\label{eq:pent}
\begin{split}
&J_{123}( 1, 1, 1, 1, 1) = \frac{1}{b}\Bigl(
a_1 J_{123}(0, 1, 1, 1, 1)+
a_2 J_{123} (1, 0, 1, 1, 1) + 
a_3 J_{123} (1, 1, 0, 1, 1) + \\
&\mspace{150mu}+a_4 J_{123}( 1, 1, 1, 0, 1)+
a_5 J_{123} (1, 1, 1, 1, 0)\Bigr)+{\cal O}(\epsilon)\,.
\end{split}
\end{equation}
This expression is obtained using the dimensional recurrence relation, where master integral $ J_{123}( 1, 1, 1, 1, 1) $ in $ d=6-2\epsilon $ dimension is expressed through a linear combination of boxes and $ J_{123}( 1, 1, 1, 1, 1) $ in $ d=4-2\epsilon $ dimension. In equation~\eqref{eq:pent} we use the following notations
\begin{equation}
\begin{split}
&a_1=-s (s_2 - t_1 + t_2-1) \bigl(3 - 3 t_1 - s_1 (2 s_1 + s_2 + t_1) + s (2 s_1 + t_1 - t_2-2) + t_2+ \\
&\mspace{100mu}+ (2 s_1 + s_2) t_2\bigr)\,,\\
&a_2=-(s_2 - t_1 + t_2-1 )\Bigl(-2 s^2 (s_1-1) + s \bigl(t_1 + s_1 (2 s_1 + 2 s_2 - t_1 - t_2) + t_2-4\bigr)-\\
&\mspace{200mu}-(1 + s_1) \bigl(s_1 (s_2 - t_1) + t_1 + t_2 - s_2 t_2-1\bigr)\Bigr)\,,\\
&a_3= s^2 (1-s_1) (2 s_2 - t_1 + t_2-2) - (1 + s_1) (s_1 + s_2-2) 
\bigl(s_1 (s_2 - t_1) + t_1 + t_2 -\\
&\mspace{50mu}- s_2 t_2-1\bigr) +s \Bigl(5 + s_2 (t_1-4) + t_1 - 3 t_2 + s_1^2 (3 s_2 - 2 t_1 + t_2-2) +\\
&\mspace{100mu}+s_1 \bigl(s_2 (2 s_2 - t_1-3) + t_1 + 2 t_2-1\bigr)\Bigr)\,,
\end{split}
\end{equation}
\begin{equation}
\begin{split}
&a_4=s^3 (t_2-t_1) + \bigl(2 (t_1-1) + (s_1 + s_2) (s_1 - t_2)\bigr) 
\bigl(s_1 (s_2 - t_1) + t_1 + t_2 - s_2 t_2-1\bigr) +\\ 
&\qquad s^2 \bigl(1 + s_2 (s_1 + t_1 - 2 t_2-2) + t_1 (3 + s_1 - t_2) + 
t_2 (t_2-1 - 2 s_1)\bigr)+\\
&\qquad+ s \Bigl(s_2^2 (t_2-s_1)-2 + 
t_1 \bigl(2 + (s_1-4) s_1 + 3 t_2\bigr) + t_2 \bigl(1 + s_1^2 - (1 + s_1) t_2\bigr) +\\
&\qquad+ s_2 \bigl(3 + 2 s_1 - 2 s_1^2 - 3 t_1 + (4 s_1 + t_1-3) t_2 - 2 t_2^2\bigr)\Bigr)\,,\\
&a_5=s \Bigl(2 t_1-2 + s_1 \bigl(1 + t_1 + s_1 (2 - s_2 + t_1)\bigr) + 
s^2 (t_1 - t_2) + t_2 -\\
&- \bigl(3 t_1 + s_1 (3 + 2 s_1 + t_1)\bigr) t_2 +(1 + 2 s_1 + s_2) t_2^2 + s \bigl(1 - 3 t_1 + t_2 + (s_2 + t_1 - t_2) t_2 + \\
&\mspace{100mu}+s_1 (s_2 - 2 t_1 + 3 t_2-2)\bigr)\Bigr)\,,
\end{split}
\end{equation}
\begin{equation}
\begin{split}
&b=2 \biggl(s^3 (s_1-1) (s_2 - t_1 + t_2-1) - 
\bigl(s_1 (s_2 - t_1) + t_1 + t_2 - s_2 t_2-1\bigr)^2 -\\
&-s^2 \bigl(3 + s_2 (t_1-3) + 2 t_1 - (3 + t_1) t_2 + t_2^2 + 
2 s_1^2 (s_2 - t_1 + t_2-1) +\\
&+ s_1 (s_2 - t_2) (s_2 - t_1 + t_2-1)\bigr) + 
s \Bigl(2 (s_2-1 ) (t_1-1) -\\
&- \bigl(2 (2 + t_1) + s_2 (s_2 + t_1-5)\bigr) t_2 + 
(2 + s_2) t_2^2 + s_1^3 (s_2 - t_1 + t_2-1)+\\
& + s_1^2 (1 + s_2 - t_2) 
(s_2 - t_1 + t_2-1) + s_1 \bigl(2 + s_2 (s_2 - t_1-3) + 4 t_1 - t_2 -\\ 
&-(1 + s_2) (s_2 - t_1) t_2 - (1 + s_2) t_2^2\bigr)\Bigr)\biggr)\,.
\end{split}
\end{equation}

All expressions for master integrals were verified numerically using the \verb|Fiesta 4| package~\cite{fiesta}.

\subsubsection*{Master-integrals for light by light corrections }

Since the number of diagrams reduces to three, we have only three families of master integrals differing by replacing the photon momenta. The first family has the following topology of integral: 
\begin{equation}\label{eq:mcb1}
\begin{split}
&I_{1}(n_1,n_2,n_3,n_4)=I_{1}(\vec{n})=\int\frac{d^{D}l}{(2\pi)^D}\frac{1}{D_1^{n_1}D_2^{n_1}D_3^{n_1}D_4^{n_1}}\\
&D_1=l^2-m^2\,,\;D_2=(l-k)^2-m^2\,,\;D_3=(l-k+k_2)^2-m^2\,,\\
&D_4=(l-k+k_1+k_2)^2-m^2\,.
\end{split}
\end{equation}
Here $ m $ is the ratio of the fermion mass ($ e,\,\mu,\,\tau $) to the mass of the muon.
In topology $ I_{1}(n_1,n_2,n_3,n_4) $ there are nine basis master integrals.
The remaining topologies are obtained by the following changes of the photon momenta  in $ I_1 $:
\begin{equation}
\begin{array}{cc}
\text{topology} & \text{changes}\\
\hline
I_{2} & k_1\leftrightarrow k_2\\
I_{3}	& k_2\leftrightarrow (-k)\\
\end{array}
\end{equation}
The basis master integrals are expressed in terms of the already introduced functions~\eqref{eq:R1}--\eqref{eq:B3}  as follows
\begin{equation}
\begin{split}
&I_1(0,0,0,1) = a_{\Gamma} m^2\Bigl(\frac{1}{\epsilon}+1-\ln(m^2)\Bigr)+{\cal O}(\epsilon),\\
&I_1(0,1,0,1) = a_{\Gamma} \biggl(\frac{1}{\epsilon}+2-\ln(m^2)+R_2\Bigl(\frac{s}{m^2}\Bigr)\biggr)+{\cal O}(\epsilon),\\
&I_1(1,0,0,1) = a_{\Gamma} \biggl(\frac{1}{\epsilon}+2-\ln(m^2)+R_2\Bigl(\frac{s-s_1-s_2+2}{m^2}\Bigr)\biggr)+{\cal O}(\epsilon),\\
&I_1(1,0,1,0) = a_{\Gamma} \biggl(\frac{1}{\epsilon}+2-\ln(m^2)+R_2\Bigl(-\frac{s_2-t_1+t_2-1}{m^2}\Bigr)\biggr)+{\cal O}(\epsilon),\\
&I_1(0,1,1,1) = \frac{a_{\Gamma}}{m^2} T_5\Bigl(\frac{s}{m^2}\Bigr)+{\cal O}(\epsilon),\\
&I_1(1,0,1,1) = \frac{a_{\Gamma}}{m^2} T_6\Bigl(\frac{s-s_1-s_2+2}{m^2},-\frac{s_2-t_1+t_2-1}{m^2}\Bigr)+{\cal O}(\epsilon),\\
&I_1(1,1,0,1) = \frac{a_{\Gamma}}{m^2} T_6\Bigl(\frac{s-s_1-s_2+2}{m^2},\frac{s}{m^2}\Bigr)+{\cal O}(\epsilon),\\
&I_1(1,1,1,0) = \frac{a_{\Gamma}}{m^2} T_5\Bigl(-\frac{s_2-t_1+t_2-1}{m^2}\Bigr)+{\cal O}(\epsilon),\\
&I_1(1,1,1,1) = \frac{a_{\Gamma}}{m^4} B_3\Bigl(\frac{s}{m^2},-\frac{s_2-t_1+t_2-1}{m^2},\frac{s-s_1-s_2+2}{m^2}\Bigr)+{\cal O}(\epsilon)\,.
\end{split}
\end{equation}

\subsubsection*{Tensor momentum integrals}

In corrections with two fermion lines there are tensor integrals in contrast to the correction with one fermion line. We introduce the following notation for the integral $ I_1[.] $ with various
integrands containing the argument of the square bracket in
the numerator, with $ l $ being the loop momentum. For instance:
\begin{equation}
I_1[(l,p_2)]=\int\frac{d^Dl}{(2\pi)^D}\frac{(l,p_2)}{D_1^{n_1}D_2^{n_2}D_3^{n_3}D_4^{n_4}}\,,
\end{equation}
where $ D_i $ are defined in~\eqref{eq:mcb1} and $ D_i $ do not contain $ p_2 $ momentum. There are the following tensor momentum integrals:
\begin{equation}\label{eq:lp2}
I_1\bigl[l^{\mu}\bigr],\;I_1\bigl[l^{\mu}l^{\nu}\bigr],\;I_1\bigl[l^{\mu}l^{\nu}l^{\alpha}\bigr],\;I_1\bigl[l^{\mu}l^{\nu}l^{\alpha}l^{\beta}\bigr]\,.
\end{equation}

Let us introduce the following denotations:
\begin{equation}
m_{ij}=(k_i,k_j)\,,\quad k_3\equiv k,\quad i=1,\,2,\,3.
\end{equation}
The metric tensor $ g^{\mu\nu}_{d-3} \equiv g_{\bot}^{\mu\nu} $ has the following properties
\begin{equation}
g^{\mu\nu}_{\bot}k_{i\mu}=0,\;g^{\mu\nu}_{\bot}g_{\mu\nu}=d-3\,.
\end{equation}
The metric tensor transverse to the vectors $ k_i $ can be represented as follows
\begin{equation}
g_{\bot}^{\mu\nu}=g^{\mu\nu}-k_i^{\mu}(m^{-1})_{ij}k_j^{\nu}\,.
\end{equation}
Here and below summation over repeated indices is implied. Let us present the projection of the vector $ l $ onto the subspace of the vectors $ k_i $ and its transverse complement
\begin{equation}
L^{\mu}= k_i^{\mu}(m^{-1})_{ij}(l,k_j)\;,\quad l_{\bot}^{\mu}=g_{\bot}^{\mu\nu}l_{\nu}.
\end{equation}
Squares of the projected vectors are:
\begin{equation}
L^2=(l,k_i)(m^{-1})_{ij}(l,k_j)\,,\quad l_{\bot}^2=l^2-L^2.
\end{equation}

Values $ L^{\mu},\,l_{\bot}^{\mu},\, L^2,\,l_{\bot}^2 $ are expressed only through scalar products $ (l,k_i),\,(k_i,k_j) $, hence master integrals with  $ L^{\mu},\,l_{\bot}^{\mu},\,L^2,\,l_{\bot}^2 $ in numerator are reduced to basis master integrals without tensors in numerator (the reduction of the numerators is done automatically with the help of the package \verb|LiteRed|). 

Tensor momentum integrals~\eqref{eq:lp2} are expressed in the following way (see Appendix in~\cite{KR:2015}):
\begin{equation}\label{eq:tensor}
\begin{split}
&I_1\bigl[l^{\mu}\bigr]=I_1\bigl[L^{\mu}\bigr]\\
&I_1\bigl[l^{\mu}l^{\nu}\bigr]=\frac{g_{\bot}^{\mu\nu}}{d-3}I_1\bigl[l_{\bot}^2\bigr]+I_1\bigl[L^{\mu}L^{\nu}\bigr]\\
&I_1\bigl[l^{\mu}l^{\nu}l^{\alpha}\bigr]=\frac{1}{d-3}I_1\bigl[\bigl(g_{\bot}^{\mu\nu}L^{\alpha}+g_{\bot}^{\mu\alpha}L^{\nu}+g_{\bot}^{\nu\alpha}L^{\mu}\bigr)l_{\bot}^2\bigr]+I_1\bigl[L^{\mu}L^{\nu}L^{\alpha}\bigr]\\
&I_1\bigl[l^{\mu}l^{\nu}l^{\alpha}l^{\beta}\bigr]=\frac{1}{(d-1)(d-3)}I_1\bigl[\bigl(g_{\bot}^{\mu\nu}g_{\bot}^{\alpha\beta}+g_{\bot}^{\mu\alpha}g_{\bot}^{\nu\beta}+g_{\bot}^{\nu\alpha}g_{\bot}^{\mu\beta}\bigr)l_{\bot}^4\bigr]+\\
&+\frac{1}{d-3}I_1\bigl[\bigl(g_{\bot}^{\mu\nu}L^{\alpha}L^{\beta}+g_{\bot}^{\mu\alpha}L^{\nu}L^{\beta}+g_{\bot}^{\nu\beta}L^{\alpha}L^{\mu}+g_{\bot}^{\alpha\beta}L^{\mu}L^{\nu}+g_{\bot}^{\mu\beta}L^{\alpha}L^{\nu}+g_{\bot}^{\nu\alpha}L^{\mu}L^{\beta}\bigr)l_{\bot}^2\bigr]+\\
&+I_1\bigl[L^{\mu}L^{\nu}L^{\alpha}L^{\beta}\bigr]\,.
\end{split}
\end{equation}

\end{document}